\documentclass[12pt]{article}
\usepackage[margin=1in]{geometry}
\usepackage{amsthm} 
\usepackage{setspace} 
\usepackage{hyperref}
\renewenvironment{abstract}{\paragraph{Abstract}}{}
\newenvironment{acknowledgements}{\paragraph{Acknowledgements}}{}
\newcommand{\keywords}[1]{\paragraph{Keywords} #1}
\newcommand{\urlB}[1]{}

\usepackage{amsmath, amssymb, amsfonts, graphicx, color, bm, float, bbm, natbib}
\newcommand{\yobs}{y_{\text{obs}}}

\newcommand{\ABC}{\text{ABC}}
\newcommand{\Nacc}{N_{\text{acc}}}

\floatstyle{ruled}
\newfloat{Algorithm}{htbp}{alg}

\title{A rare event approach to high dimensional Approximate Bayesian computation}

\author{Dennis Prangle, Richard G. Everitt and Theodore Kypraios}
\date{}

\begin{document}

\maketitle

\newcommand{\and}{, } 

\begin{abstract}
Approximate Bayesian computation (ABC) methods permit approximate inference for intractable likelihoods when it is possible to simulate from the model.
However they perform poorly for high dimensional data, and in practice must usually be used in conjunction with dimension reduction methods, resulting in a loss of accuracy which is hard to quantify or control.
We propose a new ABC method for high dimensional data based on rare event methods which we refer to as RE-ABC.
This uses a latent variable representation of the model.
For a given parameter value, we estimate the probability of the rare event that the latent variables correspond to data roughly consistent with the observations.
This is performed using sequential Monte Carlo and slice sampling to systematically search the space of latent variables.
In contrast standard ABC can be viewed as using a more naive Monte Carlo estimate.
We use our rare event probability estimator as a likelihood estimate within the pseudo-marginal Metropolis-Hastings algorithm for parameter inference.

We provide asymptotics showing that RE-ABC has a lower computational cost for high dimensional data than standard ABC methods.
We also illustrate our approach empirically, on a Gaussian distribution and an application in infectious disease modelling.
\end{abstract}



\section{Introduction} \label{sec:intro}

Approximate Bayesian computation (ABC) is a family of methods for approximate inference,
used when likelihoods are impossible or impractical to evaluate numerically
but simulating datasets from the model of interest is straightforward.
ABC can be viewed as a \emph{nearest neighbours} method.
It simulates datasets given various parameter values, and finds the closest matches, in some sense, to the observed dataset.
The corresponding parameters are used as the basis for inference.
Various Monte Carlo methods have been adapted to implement this idea, including rejection sampling \citep{Beaumont:2002}, Markov chain Monte Carlo (MCMC) \citep{Marjoram:2003} and sequential Monte Carlo (SMC) \citep{Sisson:2009}.
However it is well known that nearest neighbours approaches becomes less effective for higher dimensional data, a phenomenon referred to as the \emph{curse of dimensionality}.
The problem is that even under the best parameter values, it is rare for a high dimensional simulation to match a fixed target well, essentially because there are many random components all of which must be close matches to observations.

In this paper we propose a method to deal with this issue and permit higher dimensional data or summary statistics to be used in ABC.
The idea involves introducing latent variables $x$.
We assume data is a deterministic function $y(\theta,x)$, where $\theta$ is a vector of parameters.
Hence $x$ encapsulates all the randomness which occurs in the simulation process.
Our approach is, for a particular $\theta$ value, to use rare event methods to estimate the probability of $x$ values occurring which produce $y(\theta,x) \approx \yobs$.
As discussed later, this probability equals, up to proportionality, the approximate likelihood of $\theta$ used in existing ABC algorithms.
We estimate this probability using SMC algorithms for rare events from \cite{Cerou:2012}.
The resulting estimates are unbiased or low bias, depending on the algorithm, and can be used by many inference methods.
We concentrate on the pseudo-marginal Metropolis Hastings algorithm \citep{Andrieu:2009}, which outputs a sample from a distribution approximating the Bayesian posterior.

The intuition for the rare event probability estimates we use is as follows.
Given $\theta$, standard ABC methods effectively simulate one or several $x$ values from their prior and calculate a Monte Carlo estimate of $\Pr(y(\theta,x) \approx \yobs)$.
This relative error of this estimate has high variance when the probability is small, as is the case when we require close matches.
The rare event technique of \emph{splitting} uses nested sets of latent variables $A_1 \supset A_2 \supset \ldots \supset A_T$, representing increasingly close matches.
We aim to estimate $\Pr(A_1)$, $\Pr(A_2 | A_1)$, $\Pr(A_3 | A_2), \ldots$ and take the product.
If these probabilities are all relatively large then the variance of the final estimator's relative error is smaller than using a single stage of Monte Carlo
(For a crude variance analysis justifying this see \citealp{L'Ecuyer:2007}.
\citealp{Cerou:2012} prove more detailed results for their SMC algorithms which we summarise later.)
We can estimate $\Pr(A_1)$ using Monte Carlo with $N$ samples.
Next we reuse the $x$ samples with $x \in A_1$.
We sample randomly from these $N$ times and, to avoid duplicates, perturb each appropriately.
We found a good perturbation method was a slice sampling algorithm from \cite{Murray:2016}.
The resulting sample is used to find a Monte Carlo estimate of $\Pr(A_2 | A_1)$.
We carry on similarly to estimate the remaining conditional probabilities.

For this approach to work well, a small perturbation of the $x$s must produce a corresponding small perturbation of the $y$s.
Hence the mapping $y(\theta,x)$ must be well-chosen.
This requirement is explored in Section \ref{sec:latent}.

We consider two rare event SMC algorithms proposed by \cite{Cerou:2012}.
In one the nested sets must be fixed in advance
and in the other they are selected adaptively during the algorithm.
A contribution of this paper is to compare the efficiency of these algorithms within the setting of ABC.
Our recommendation, discussed in Section \ref{sec:adaptation}, is a combination of the two approaches:
a single run of the adaptive algorithm to select the nested sets,
followed by using these in the fixed algorithm.

\subsection{Related literature}

First we highlight the difference between our approach and ABC-SMC \citep{Sisson:2009, DelMoral:2012}.
These methods find parameter values which are most likely to produce simulations closely matching the observations.
We argue that for high dimensional observations, such simulations are rare even for the best parameter values.
Instead we use SMC in a different way, to find latent variables which produce successful simulations.
In Section \ref{sec:extensions} we discuss the possibility of combining these two approaches.
Another method that seeks to find promising parameter values is ABC subset simulation \citep{Chiachio:2014}.
To our knowledge this is the only other approach to ABC using rare event methods.
Again, our approach differs from this by instead searching a space of latent variables.

The most popular approach to deal with the curse of dimensionality in ABC is \emph{dimension reduction}.
Here high-dimensional datasets are mapped to lower dimensional vectors of features, often referred to as \emph{summary statistics}.
The quality of a match between simulated and observed data is then judged based only on their corresponding summary vectors.
However, using summary statistics involves some loss of information about the posterior which is hard to quantify.
Low dimensional sufficient statistics would avoid this problem but generally do not exist,
and there are many competing methods to choose summaries which make a good trade-off between low dimension and informativeness \citep{Blum:2013, Prangle:2017}.
An alternative approach of \cite{Nott:2014} is to improve ABC output by adjusting each parameter's margin to agree with a separate marginal ABC analysis.
These analyses can each use different low dimension summary statistics, so that the effect of the curse of dimensionality on the margins is reduced.
However there are still issues in selecting these summaries, and dealing with approximation error in the dependence structure.
Recently, an extension has looked at assuming a Gaussian copula dependence structure \citep{Li:2017}.
More high dimensional ABC methods are reviewed in \cite{Nott:2017}.

Several other authors have recently investigated latent variable approaches to ABC.
\cite{Neal:2012} introduced \emph{coupled ABC} for household epidemics.
This simulates latent variable vectors from their prior and, for each, finds one or many parameter vectors leading to closely matching simulated datasets.
These parameters, weighted appropriately, form a sample from an approximate posterior.
A similar strategy is employed for more general applications in \cite{Meeds:2015}'s \emph{optimisation Monte Carlo} and the \emph{reverse sampler} of \cite{Forneron:2016}.
Alternatively, \cite{Moreno:2016} perform variational inference, using latent variable vectors drawn from their prior in the estimation of loss function gradients.
Another related method is \cite{Graham:2016}, who sample from the $(\theta,x)$ space conditioned exactly on the observations using constrained Hamiltonian Monte Carlo (HMC).
A limitation is that the $y(\theta,x)$ mapping must be differentiable with respect to both arguments.

A similar SMC approach to ours is outlined, but not implemented, by \cite{Andrieu:2012}.
Analogous methods have been implemented for ABC inference of state space models, using ABC particle filtering to estimate likelihoods for a sequence of observations \citep{Jasra:2015}.

\cite{Targino:2015} use similar methods to us in a non-ABC context.
They use SMC to estimate posterior quantities for a copula model conditional on a rare event.
Like us, they use increasingly rare events as intermediate targets, and use slice sampling for perturbation moves.
A difference is our focus on estimating the probability of the rare event, and providing results on the asymptotic efficiency of this.
Also their perturbation updates each component of $x$ in turn with a univariate slice sampler, while we use truly multivariate updates.

\subsection{Contributions and overview}

We provide an approximate inference method for the same class of intractable problems as ABC.
Our algorithm samples from the same family of posterior approximations as ABC, but can reach more accurate approximations for the same computational cost.
In particular, its cost rises more slowly with the data dimension.
Therefore it is feasible to perform inference using a larger, and hence more informative, set of summary statistics.
In some cases it is even feasible to use the full data.

Our method has various differences to competing methods using latent variables.
Unlike the majority of these, it does not rely solely on randomly sampling latent variables, but instead searches their space more efficiently.
Also unlike HMC approaches we do not require differentiability assumptions for $y(\theta,x)$.

Typically SMC methods have many tuning choices.
Another benefit of our approach is that these can all be automated.
The tuning choices required are simply those for the ABC and PMMH algorithms.

Section \ref{sec:background} describes background information on the methods we use.
Section \ref{sec:hdABC} presents our algorithm to estimate the likelihood given a particular parameter vector, and how we use this within a MCMC inference algorithm.
Asymptotic results on computational cost are also given here, quantifying the improvement over standard ABC.
The method is evaluated on a simple Gaussian example in Section \ref{sec:Gaussian}, and used in an infectious disease application in Section \ref{sec:epidemics}.
Code for these examples is available at
\url{https://github.com/dennisprangle/RareEventABC.jl}\urlB{https://github.com/dennis\\}\urlB{prangle/RareEventABC.jl}. 
Section \ref{sec:conclusion} gives a concluding discussion, including when we expect our scheme to work well.
Appendix \ref{sec:cost} contains technical details of our asymptotics.

\section{Background} \label{sec:background}

\subsection{Approximate Bayesian Computation} \label{sec:ABC}

Suppose observations $\yobs$ are available and we wish to learn the parameters $\theta$ of a model $\pi(y | \theta)$ (a density with respect to a probability measure $dy$) given a prior $\pi(\theta)$ (a density with respect to probability measure $d\theta$).
Algorithm \ref{alg:ABCrs} is an ABC rejection sampling algorithm which performs approximate Bayesian inference.
It requires three tuning choices:
the number of simulations $N$,
a threshold $\epsilon \geq 0$,
and a distance function $d(\cdot,\cdot)$.
The latter is typically Euclidean distance or a variation. 
It is usually sensible to scale data $y$ appropriately so that all components make contributions of similar size to the distance, and we will assume that this has already been done.

\begin{Algorithm}[htp]
\caption{ABC rejection sampling} \label{alg:ABCrs}
\begin{enumerate}
\item[] {\bf Loop over $i=1,2,\ldots,N$}.
\item Sample $\theta_i$ from $\pi(\theta)$.
\item Sample $y_i$ from $\pi(y|\theta_i)$.
\item Accept if $d(y_i, \yobs) \leq \epsilon$. \label{step:accept}
\item[] {\bf End loop}
\item {\bf Return}: accepted $\theta_i$ values.
\end{enumerate}
\end{Algorithm}

The output of Algorithm \ref{alg:ABCrs} is a sample from the following approximate posterior density
\begin{equation}
\pi_\ABC(\theta | \yobs; \epsilon) \propto \pi(\theta) L_\ABC(\theta; \epsilon) \label{eq:ABCposterior},
\end{equation}
where
\begin{align}
L_\ABC(\theta; \epsilon) &= V(\epsilon)^{-1} \int \pi(y|\theta) \mathbbm{1}[d(y, \yobs) \leq \epsilon] dy, \label{eq:ABClikelihood} \\
V(\epsilon) &= \int \mathbbm{1}[d(y, \yobs) \leq \epsilon] dy. \label{eq:V}
\end{align}
The \emph{ABC likelihood} $L_\ABC$ is a convolution of the exact likelihood function and the \emph{kernel}
\begin{equation} \label{eq:uniform kernel}
k(y; \epsilon) = V(\epsilon)^{-1} \mathbbm{1}[d(y, \yobs) \leq \epsilon],
\end{equation}
a uniform density on $y$ values close to $\yobs$.
Under some weak conditions, as $\epsilon \to 0$ the ABC likelihood converges to the exact likelihood and $\pi_\ABC$ to the exact posterior
(This is shown by equation \eqref{eq:LDT} in Appendix \ref{sec:cost}, which describes some sufficient conditions.)
However this causes acceptances to become rare.
Thus there is a trade-off, controlled by $\epsilon$, between output sample size and the accuracy of $\pi_\ABC$.


ABC rejection sampling is inefficient in the common situation where the prior is much more diffuse than the posterior, as a lot of time is spent on simulations that have very little chance of being accepted.
Several more sophisticated ABC algorithms have been proposed which concentrate on performing simulations for $\theta$ values believed to have high posterior density.
These include versions of importance sampling, MCMC 
and SMC. 
These also output samples (sometimes weighted) from an approximation to the posterior, usually $\pi_\ABC$ as given in \eqref{eq:ABCposterior}.
See \cite{Marin:2012} for a review of ABC, including these algorithms and related theory.

As mentioned earlier, ABC suffers from a \emph{curse of dimensionality} issue.
Intuitively, the problem is that simulations producing good matches of all summaries simultaneously become increasingly unlikely as $\dim(y)$ grows.
For Algorithm \ref{alg:ABCrs}, it has been proved \citep{Blum:2010, Barber:2015, Biau:2015}
that for a fixed value of $N$ the quality of the output sample as an approximation of the posterior deteriorates as $d$ increases, even taking into account the possibility of adjusting $\epsilon$.
See \cite{Fearnhead:2012} for heuristic arguments that the problem also applies to other ABC algorithms.

\subsection{Pseudo-marginal Metropolis-Hastings} \label{sec:PMMH}

The approach of this paper is to estimate the ABC likelihood \eqref{eq:ABClikelihood} more accurately than standard ABC methods.
This section reviews one approach for how such estimates can be used to sample from $\pi_\ABC$.

The Metropolis-Hastings (MH) algorithm samples from a Markov chain with stationary distribution proportional to an unnormalised density $\psi(\theta)$.
It is often used in Bayesian inference to produce samples from a close approximation to the posterior distribution.
Despite the non-independence of these samples, they can still be used to produce highly accurate Monte Carlo estimates of functions of the posterior.
Simulating $\theta_t$, the $t$th state of the Markov chain is based on sampling a state $\theta'$ from a proposal density $q(\theta'|\theta_{t-1})$, typically centred on the preceding state $\theta_{t-1}$.
This proposal is accepted as $\theta_t$ with probability
$
\min \left(1,
\frac{\psi(\theta') q(\theta_{t-1}|\theta')}{\psi(\theta_t) q(\theta'|\theta_{t-1})}
\right)
$.
Otherwise $\theta_t=\theta_{t-1}$.

This algorithm remains valid if likelihood evaluations are replaced with unbiased non-negative estimates as follows \citep{Andrieu:2009}.
The state of the Markov chain is now $(\theta_t, \hat{\psi}_t)$, where $\hat{\psi}_t$ is an estimate of $\psi(\theta_t)$, and the acceptance probability must be
$
\min \left(1,
\frac{\hat{\psi}' q(\theta_{t-1}|\theta')}{\hat{\psi}_{t-1} q(\theta'|\theta_{t-1})}
\right).
$
Crucially, upon acceptance $\hat{\psi}_t$ is set to the estimate $\hat{\psi}'$ for the proposal $\theta'$.
So, rather than being recalculated in every iteration, this estimate is used in all future iterations until another proposal is accepted.
A version of the resulting \emph{pseudo-marginal Metropolis-Hastings} (PMMH) algorithm, specialised to this paper's setting, is presented below as Algorithm \ref{alg:pmmh}.

Optimal tuning of PMMH has been examined theoretically by \cite{Pitt:2012}, \cite{Doucet:2015} and \cite{Sherlock:2015}, covering the case where each $\hat{\psi}'$ estimate is generated by an SMC algorithm.
A central issue is how many SMC particles should be used to optimise the computational efficiency of PMMH.
All the authors conclude that this number should be tuned to achieve a particular variance of $\log \hat{\psi}$.
(It's assumed, unrealistically, that this variance does not depend on $\theta$.
In practice it's typical to investigate the variance at a fixed value of $\theta$ believed to have high posterior density.)
The value derived for this optimal variance differs between the authors due to their different assumptions, but all values lie in the range 0.8--3.3.
\cite{Sherlock:2015} also investigate tuning the proposal distribution $q$, and suggest using proposal variance $\frac{2.562^2}{\dim(\theta)} \Sigma$ where $\Sigma$ is the posterior variance.
They perform simulation studies generally supporting both these results.
One key assumption made by all the authors is that $\log \hat{\psi}$ follows a normal distribution.
The validity of this assumption in our setting will be investigated later.
It's also assumed that the computational cost of SMC is proportional to the number of particles used and does not depend on $\theta$, which is generally true for SMC algorithms.


\subsection{Rare event sequential Monte Carlo} \label{sec:RE-SMC}

To estimate the ABC likelihood \eqref{eq:ABClikelihood} in Section \ref{sec:hdABC} we will use two algorithms of \cite{Cerou:2012} for estimating rare event probabilities using a SMC approach.
This section reviews existing work on these algorithms.
A few novel remarks which are relevant later are given at the end.

The aim is to estimate a small probability, $P=\Pr(\Phi(x) \leq \epsilon | \theta)$.
Here $x$ is a random variable, $\theta$ is a vector of parameters, $\Phi$ maps $x$ values to $\mathbb{R}$, and $\epsilon$ is a threshold.
In the ABC setting of later sections $P$ will be an estimate of $L_\ABC(\theta;\epsilon)$ up to proportionality.
As discussed informally in Section \ref{sec:intro}, both algorithms act by estimating conditional probabilities $\Pr(\Phi(x) \leq \epsilon_{k+1} | \theta, \Phi(x) \leq \epsilon_k)$ for a decreasing sequence of $\epsilon$ values.
In Algorithm \ref{alg:FIXED-RE-SMC} (FIXED-RE-SMC) a fixed $\epsilon$ sequence must be prespecified.
In Algorithm \ref{alg:ADAPT-RE-SMC} (ADAPT-RE-SMC) the sequence is selected adaptively.
Whenever we use RE-SMC without an additional prefix we are referring to both algorithms.

\cite{Cerou:2012} prove that FIXED-RE-SMC produces an unbiased estimator of $P$, but ADAPT-RE-SMC gives an estimator with $O(N^{-1})$ bias.
They also analyse the asymptotic variance of the estimators' relative errors for large $N$ under various assumptions.
This variance is generally smaller for ADAPT-RE-SMC.
Equality occurs only when FIXED-RE-SMC uses an $\epsilon$ sequence such that $\Pr(\Phi(x) \leq \epsilon_{k+1} | \theta, \Phi(x) \leq \epsilon_k)$ is constant as $k$ varies.
An approximation to this sequence can be generated by running ADAPT-RE-SMC.
We discuss which RE-SMC algorithm to use within our method later.
Under optimal conditions the relative error variances decrease as $T$, the number of iterations, grows, so that the estimates are more accurate than using plain Monte Carlo, which corresponds to $T=1$.
This result could be extended to take computational cost into account.
However instead we will analyse the overall efficiency of our proposed approach in Section \ref{sec:cost summary}.

\begin{Algorithm}[hpt]
\caption{Rare event SMC algorithm, with fixed $\epsilon$ sequence (FIXED-RE-SMC)} \label{alg:FIXED-RE-SMC}
\begin{itemize}
\item[] {\bf Input}: Parameters $\theta$, number of particles $N$, thresholds $\epsilon_1,\epsilon_2,\ldots,\epsilon_T$, Markov kernels for step 3.
\end{itemize}
\begin{enumerate}
\item
For $i = 1,\ldots, N$ sample $x_0^{(i)}$ from $\pi(x|\theta)$.
\item[]
\textbf{Loop over $1 \leq t \leq T$:}
\item
Calculate $I_t = \{ i | \Phi(x_{t-1}^{(i)}) \leq \epsilon_t \}$.
Let $\hat{P}_t = |I_t| / N$.\\
(If $\hat{P}_t=0$ terminate algorithm returning $\hat{P}=0$.)
\item
For $i=1,\ldots,N$ sample $x^{(i)}_t$ by drawing $j$ uniformly from $I_t$ and applying a Markov kernel to $x^{(j)}_{t-1}$
with invariant density $\pi(x|\theta,\Phi(x) \leq \epsilon_{t-1})$ (taking $\epsilon_0 = \infty$).
\item[]
\textbf{End loop}
\item
{\bf Return}: $\hat{P} = \prod_{t=1}^T \hat{P}_t$. 
\end{enumerate}
\end{Algorithm}

\begin{Algorithm}[hpt]
\caption{Rare event SMC algorithm, with adaptive $\epsilon$ sequence (ADAPT-RE-SMC)} \label{alg:ADAPT-RE-SMC}
\begin{itemize}
\item[] {\bf Input}: Parameters $\theta$, number of particles $N$, target number to accept $\Nacc$, acceptance thresholds $\epsilon$, rule to generate Markov kernels for step 3.
\end{itemize}
\begin{enumerate}
\item
For $i = 1,\ldots, N$ sample $x_0^{(i)}$ from $\pi(x|\theta)$.
\item[]
\textbf{Loop over $t=1,2,\ldots$:}
\item
Let $\epsilon_t$ be the maximum of (a) the $\Nacc$th smallest $\Phi(x^{(i)}_{t-1})$ value and (b) $\epsilon$. \\
Calculate $I_t = \{ i | \Phi(x_{t-1}^{(i)}) \leq \epsilon_t \}$ and $\hat{P}_t = |I_t| / N$.
\item
For $i=1,\ldots,N$ sample $x^{(i)}_t$ by drawing $j$ uniformly from $I_t$ and applying a Markov kernel to $x^{(j)}_{t-1}$ with invariant density $\pi(x|\theta,\Phi(x) \leq \epsilon_{t-1})$ (taking $\epsilon_0 = \infty$).
\item
If $\epsilon_t = \epsilon$ break loop and go to step 5, setting $T=t$.
\item[]
\textbf{End loop}
\item
{\bf Return}: $\hat{P} = \prod_{t=1}^T \hat{P}_t$.
\end{enumerate}
\end{Algorithm}

\paragraph{Remarks}
\begin{enumerate}
\item Step 2 of ADAPT-RE-SMC selects a threshold sequence in the same way as the ABC-SMC algorithm of \cite{DelMoral:2012}.
Unlike that work however, this sequence is specialised to one particular $\theta$ value rather than being used for many proposed $\theta$s.
\item In ADAPT-RE-SMC, typically $\Nacc$ particles are accepted so that $|I_t|=\Nacc$.
However there may be more acceptances in the final iteration or if ties in distance are possible.
\item For $t \leq T$, $\prod_{\tau=1}^t \hat{P}_\tau$ is an upper bound on $\hat{P}$ in either RE-SMC algorithm.
This bound can be calculated during the $t$th iteration of the algorithms.
This will be used below to terminate the algorithms early once the estimate is guaranteed to be below some prespecified bound.
\item The $x^{(i)}_T$ values can be used for inference of $x | \theta, \Phi(x) \leq \epsilon$.
When this is not of interest, as in this paper, then the computational cost can be reduced by omitting step 3 (resampling and Markov kernel propagation) in the final iteration of either algorithm.
\item It's possible for ADAPT-RE-SMC not to terminate.
This could occur if the $x^{(i)}_t$ particles become stuck near a mode where $\Phi(x) > \epsilon$ and the Markov kernel is unable to move them to other modes.
In Section \ref{sec:inference} we will discuss how our proposed method can avoid this problem by terminating once it becomes clear the final likelihood estimate will be very low.
\item When ties in the distance are possible, ADAPT-RE-SMC iterations can fail to reduce the threshold.
That is, sometimes step 2 can give $\epsilon_{t+1}=\epsilon_t$.
This can produce very long run times.
Possible improvements to deal with this are discussed in Section \ref{sec:extensions}.
(Note that when ADAPT-RE-SMC is being used to select a sequence of thresholds then repeated values should be removed.)
\item These algorithms use multinomial resampling.
More efficient schemes exist, but are not investigated by the theoretical results of \cite{Cerou:2012}.
\end{enumerate}

\subsection{Slice sampling} \label{sec:slice}


We require a suitable Markov kernel to use within the RE-SMC algorithms.
This must have invariant density $\pi(x|\theta, \Phi(x) \leq \epsilon_{t-1})$.
As discussed below in Section \ref{sec:inference}, our ABC setting will assume $\pi(x|\theta)$ is uniform on $[0,1]^m$.
Hence the required invariant distribution is uniform on the subset of $[0,1]^m$ such that $\Phi(x) \leq \epsilon_{t-1}$.
We will use \emph{slice sampling} as the Markov kernel.
This section outlines the general idea of slice sampling and a particular algorithm.
We also include some novel material on how it can be adapted to our setting and advantages over alternative choices.

Slice sampling is a family of MCMC methods to sample from an unnormalised target density $\gamma(x)$.
The general idea is to sample uniformly from the set $\{ (x,h)$ $| h \leq \gamma(x) \}$ and marginalise.
We will concentrate on an algorithm of \cite{Murray:2016} for the case where the support of $\gamma(x)$ is $[0,1]^m$, or a subset of this.
Their algorithm updates the current state $x$ by first drawing $h$ from $\text{Uniform}(0,\gamma(x))$,
then proposing $x'$ values, accepting the first one for which $\gamma(x') \geq h$.
The proposal scheme initially considers large changes from $x$ in a randomly chosen direction, and then, if these are rejected, progressively smaller changes.

For use within RE-SMC, $\gamma(x)$ can be taken to be the indicator function $\mathbbm{1}(\Phi(x) \leq \epsilon_{t-1})$.
This means the condition $\gamma(x') \geq h$ simplifies to $\gamma(x') > 0$, so sampling $h$ can be omitted.
The resulting slice sampling update is given by Algorithm \ref{alg:slice}, which is a special case of the \cite{Murray:2016} algorithm mentioned above (and similar to the \emph{hit-and-run sampler}; see \citealp{Smith:1996}).
See their paper for details of the proof that $\gamma(x)$ is the invariant density of this Markov kernel.

\begin{Algorithm}[htp]
\caption{Slice sampling update for RE-SMC} \label{alg:slice}
\begin{itemize}
\item[] {\bf Input}: current state $x$ of dimension $p$, map $\Phi(x)$, threshold $\epsilon$, initial search width $w$. It's assumed that $\Phi(x) \leq \epsilon$.
\end{itemize}
\begin{enumerate}
\item Sample $v \sim N(0,I_p)$
\item Sample $u \sim \text{Uniform}(0,w)$. Let $a=-u, b=w-u$.
\item[] {\bf Loop:}
\item Sample $z \sim \text{Uniform}(a,b)$.
\item Define a vector $x'$ by $x'_i=r(x_i+zv_i)$ using the \emph{reflection function}:
\[
r(y) =
\begin{cases}
m   & m<1      \\
2-m & m \geq 1
\end{cases}
\]
where $m$ is the remainder of $y$ modulo 2.
\item If $\Phi(x') \leq \epsilon$ then {\bf return} $x'$.
\item If $z<0$ let $a=z$, otherwise let $b=z$.
\item[] {\bf End loop}
\end{enumerate}
\end{Algorithm}

Next we describe two advantages of using slice sampling within RE-SMC, particularly in relation to the alternative of using a Metropolis-Hastings kernel.
Firstly, slice sampling requires little tuning.
If tuning choices were required, for example a proposal distribution for Metropolis-Hastings, then RE-SMC would need to include rules to make a good choice automatically, which may be difficult.
Another advantage of slice sampling is that each iteration outputs a unique $x$ value.
On the other hand, Metropolis-Hastings rejections can lead to duplicates, which is problematic within SMC because it leads to increased variance of probability estimates.

The only tuning choice required by Algorithm \ref{alg:slice} is the initial search width $w$.
A default choice is $w=1$, but this means that the number of loops required will increase for small $\epsilon$.
To deal with this we choose $w=1$ in the first SMC iteration and then select $w$ adaptively, as $\min(1, 2 \bar{z})$ where $\bar{z}$ is the maximum final value of $|z|$ from all slice sampling calls in the previous SMC iteration.
This choice generally shrinks $w$ based on the most recent value of $\bar{z}$, while avoiding some unwanted behaviours.
Firstly it avoids forcing $w$ to decrease at a fixed rate, so that eventually only very small steps would be attempted.
Secondly it avoids $w$ growing above 1, which would make slice sampling expensive when local moves are required.
The effect of our choice is investigated empirically later (see Figure \ref{fig:slice}).


\section{High dimensional ABC} \label{sec:hdABC}

This section presents our approach to inference in the ABC setting, using the algorithms reviewed in Section \ref{sec:background}.
Section \ref{sec:Lest} describes how the RE-SMC algorithms can estimate the ABC likelihood given values of $\theta$ and $\epsilon$, and a latent variable structure.
Such likelihood estimators can be used within several inference algorithms to produce approximate Bayesian inference.
In this paper we concentrate on PMMH.
Section \ref{sec:inference} presents the resulting method.
Section \ref{sec:cost summary} discusses the computational cost of the resulting \emph{RE-ABC algorithm} in comparison to standard ABC, with particular note of the high dimensional case.

Two versions of RE-ABC are possible, depending on whether likelihood estimates are produced using FIXED-RE-SMC or ADAPT-RE-SMC.
We present both and compare them throughout the remainder of the paper.
As will be explained, in Sections \ref{sec:inference} and \ref{sec:adaptation}, we conclude by arguing in favour of using FIXED-RE-SMC together with an initial run of ADAPT-RE-SMC to select the $\epsilon$ sequence.

\subsection{Likelihood estimation} \label{sec:Lest}

For now, suppose $\theta$ and $\epsilon>0$ are fixed.
We aim to produce an unbiased estimate of $L_\ABC(\theta; \epsilon)$, as defined in \eqref{eq:ABClikelihood}.

Suppose there exist latent variables $x$ such that the observations can be written as a deterministic function $y=y(\theta, x)$.
The idea is that $x$ and $\theta$ suffice to specify a complete realisation of the simulation process, even including details such as observation error, and $y(\theta, x)$ is a vector of partial observations.
Neglecting $\theta$, which is fixed for now, $y(\theta, x)$ will be written below as simply $y=y(x)$.
See Section \ref{sec:latent} for a discussion of properties of $y(\theta, x)$ which help our approach work well.

We specify a density $\pi(x|\theta)$ (with respect to Lebesgue measure) for the latent variables.
This is part of the specification of the model, but it can also be viewed as representing prior beliefs about the latent variables.
Throughout the paper we take $\pi(x|\theta)$ to be uniform on $[0,1]^m$ regardless of $\theta$.
Under this interpretation, $x$ is a vector of $m$ independent standard uniform random variables which suffice to carry out the simulation process.

Now we simply apply one of the RE-SMC algorithms using $\Phi(x) = d(y(x), \yobs)$.
The small probability estimated by these algorithms is
\begin{align*}
\Pr(\Phi(x) \leq \epsilon | \theta)
&= \int \pi(x|\theta) \mathbbm{1}[d(y(x), \yobs) \leq \epsilon_t] dx \\
&= \int \pi(y|\theta) \mathbbm{1}[d(y, \yobs) \leq \epsilon_t] dy,
\end{align*}
which equals $L_\ABC(\theta; \epsilon)$ multiplied by the constant $V(\epsilon)$.
Hence using FIXED-RE-SMC we can obtain an estimate of $L_\ABC(\theta; \epsilon)$ which is unbiased, as required by PMMH.
Using ADAPT-RE-SMC produces a slightly biased estimate, and we comment on the effect of using this within PMMH in the next section.

Note that we assume $\pi(x|\theta)$ to be uniform simply for convenience.
Firstly, many latent variable representations can easily be re-expressed in this form.
Secondly, given this assumption, the slice sampling method of Algorithm \ref{alg:slice} is well-suited to be the Markov kernel within RE-SMC.
Our methodology could be adapted to use other $\pi(x|\theta)$ distributions if desired.
The main change needed would be to use alternative Markov kernels, for example elliptical slice sampling (see \citealp{Murray:2016}) for the Gaussian case, or Gibbs updates for the discrete case.
These changes could well improve performance for particular applications.

\subsection{Inference} \label{sec:inference}

Algorithm \ref{alg:pmmh} shows the PMMH algorithm for our setting, which we refer to as RE-ABC.
It can use either FIXED-RE-SMC or ADAPT-RE-SMC when estimates of the ABC likelihood are required.
We'll use the prefixes FIXED and ADAPT to refer to the version of RE-ABC based on the corresponding RE-SMC algorithm.

\begin{Algorithm}[htp]
\caption{Pseudo-marginal Metropolis-Hastings using RE-SMC (RE-ABC)} \label{alg:pmmh}
\label{alg:effIS}
\begin{enumerate}
\item[] {\bf Input}: initial state $\theta_1$, number of iterations $M$, number of SMC particles $N$ and tuning choices for PMMH, RE-SMC and ABC.
\item Let $t=1$ and calculate $\hat{L}_{\ABC, 1}$ using FIXED-RE-SMC (or ADAPT-RE-SMC) with slice sampling as the Markov kernel.
\item[] {\bf Loop over $2 \leq t \leq M$}:
\item Propose new state $\theta'$ from $q(\cdot|\theta)$ and sample $u$ from $\text{Uniform}(0,1)$.
\item
Let $\hat{L}_{\ABC}'$ be the output of FIXED-RE-SMC (or ADAPT-RE-SMC) with slice sampling as the Markov kernel.
This algorithm can be stopped early once rejection in the next step is guaranteed.
\item \label{step:PMMHacc}
If $u > \frac{\pi(\theta') \hat{L}_{\ABC}' q(\theta_{t-1}|\theta')}{\pi(\theta_{t-1}) \hat{L}_{\ABC, t-1} q(\theta'|\theta_{t-1})}$:
\begin{itemize}
  \item[] \emph{Reject} Let $\theta_t = \theta_{t-1}$ and $\hat{L}_{\ABC, t} = \hat{L}_{\ABC, t-1}$.
\end{itemize}
Else:
\begin{itemize}
  \item[] \emph{Accept} Let $\theta_t = \theta'$ and $\hat{L}_{\ABC, t} = \hat{L}_{\ABC}'$.
\end{itemize}
\item[] {\bf End loop}
\item[] {\bf Return}: $\theta_1, \theta_2, \ldots, \theta_M$.
\end{enumerate}
\end{Algorithm}

For FIXED-RE-ABC, the likelihood estimates are unbiased estimates of $L_{\ABC}(\theta; \epsilon)$ up to proportionality.
Therefore the probability of acceptance in step \ref{step:PMMHacc} corresponds to a target density proportional to $\pi(\theta) L_\ABC(\theta; \epsilon)$ i.e.~the standard ABC posterior \eqref{eq:ABCposterior}.
ADAPT-RE-ABC involves biased likelihood estimates so does not sample from exactly this density.
However the bias introduced is small, and may have little effect compared to the efficiency benefits of the variance reduction which ADAPT-RE-SMC provides (theoretical and practical aspects of MCMC algorithms that have this character are discussed in \citealt{Alquier:2016}).
We investigate this empirically in Sections \ref{sec:Gaussian} and \ref{sec:epidemics} and find no noticeable effect of bias.
However we find ADAPT-RE-SMC to sometimes be less computationally efficient in practice, and so we recommend using the FIXED-RE-SMC algorithm, together with a single run of ADAPT-RE-SMC to select a $\epsilon$ sequence.
Reasons for this are described shortly, and discussed in more detail in Section \ref{sec:conclusion},
together with possibilities for improvement.

To reduce computational costs RE-SMC can be terminated as soon as rejection is guaranteed.
To implement this, after step 2 of RE-SMC check whether
\[
\prod_{\tau=1}^{t_{\text{SMC}}} \hat{P}_{\tau} < \frac{u \pi(\theta_{t-1}) \hat{L}_{\ABC, t-1} q(\theta'|\theta_{t-1})}{\pi(\theta') q(\theta_{t-1}|\theta')},
\]
where $t_{\text{SMC}}$ is the current value of the $t$ variable within the RE-SMC algorithm.
If this is true, terminate the RE-SMC algorithm and reject the current proposal in the PMMH algorithm.
The MCMC algorithm remains valid since the final RE-SMC likelihood estimate is guaranteed to be smaller than $\prod_{\tau=1}^{t_{\text{SMC}}} \hat{P}_{\tau}$ and therefore lead to rejection in PMMH.
Early termination prevents extremely long runs of RE-SMC for $\theta$ values with low posterior densities.
It is most efficient for FIXED-RE-SMC, where it is always possible to terminate in any iteration if the $\hat{P}_{\tau}$ values are small enough.
For ADAPT-RE-SMC, $\hat{P}_{\tau} \geq \frac{\Nacc}{N}$ so there is a lower bound of how many iterations are required before termination.
This argument suggests ADAPT-RE-SMC is less computationally efficient, and agrees with later empirical findings (see Figure \ref{fig:times_Abakaliki}).

Earlier we commented that ADAPT-RE-SMC can fail to terminate in some situations.
When ties in the distance are not possible, then this is usually not a problem within RE-ABC due to the early termination rule just outlined.
However care is still required the first time ADAPT-RE-SMC is run, and when it is used in pilot runs.
Ties in the distance are potentially more problematic and are discussed further in Section \ref{sec:extensions}.

There are numerous tuning choices required in this PMMH algorithm.
Most of these can be based on the output of a pilot analysis,
for example an ABC analysis or a short initial run of PMMH.
The estimated posterior mean $\hat{\mu}$ can be used as an initial PMMH state.
The estimated posterior variance $\hat{\Sigma}$ can be used to tune the PMMH proposal density.
Following the PMMH theory discussed in Section \ref{sec:PMMH} we sample proposal increments from $N \left( 0, \frac{2.562^2}{\dim(\theta)} \hat{\Sigma} \right)$.
(Note that the early termination rule avoids SMC calls having very long run times for some $\theta$ values, approximately meeting the assumptions of the PMMH tuning literature.)
The threshold sequence for FIXED-RE-SMC can be selected by running ADAPT-RE-SMC with $\theta=\hat{\mu}$.
To select the number of particles, a few preliminary runs of FIXED-RE-SMC (or ADAPT-RE-SMC) can be performed with $\theta=\hat{\mu}$, aiming to produce a log likelihood variance of roughly $1$.
This is at the more conservative end of the range suggested by the theory reviewed earlier.

A crucial tuning choice which remains is $\epsilon$.
As in other ABC methods, we suggest tuning this pragmatically based on the computational resources available.
This can be done by running ADAPT-RE-SMC with $\theta=\hat{\mu}$ and $\epsilon=0$ and stopping after a prespecified time, corresponding to how long is available for an iteration of PMMH.
The value of $\epsilon_t$ when the algorithm is stopped can be used as $\epsilon$.
It is still possible for the SMC algorithms to take much longer to run for other $\theta$ values.
However the early termination rule will usually mitigate this.
Diagnostic plots can be used to investigate whether the $\epsilon$ value selected produces simulations judged to be sufficiently similar to the observations.
For example, see Figure 1 of the supplementary material.

\subsection{Cost} \label{sec:cost summary}

Here we summarise results on the cost of ABC and RE-ABC in terms of time per samples produced (or effective sample size for PMMH algorithms), in the asymptotic case of small $\epsilon$.
Arguments supporting these results are given in Appendix \ref{sec:cost}.
Several assumptions are required, principally that $\pi(y|\theta)$ is a density with respect to Lebesgue measure -- informally, the observations must be continuous.
Weakening these assumptions is discussed in supplementary material.
Note that the results are the same whether FIXED-RE-ABC and ADAPT-RE-ABC is used.

The time per sample is asymptotic to $1/V(\epsilon)$ for ABC and $[\log V(\epsilon)]^2$ for RE-ABC
(see \eqref{eq:V} for definition of $V(\epsilon)$.)
So, asymptotically, RE-ABC has a significantly lower cost to reach the same target density.
To illustrate the effect of $D = \dim(y)$ we can consider the asymptotic case of large $D$
(n.b.~as shown in the supplementary material, when some observations are non-continuous then $D$ can be replaced with the dimension of $\{ y | d(y,\yobs) < \epsilon\}$ for small $\epsilon$.)
Under the Lebesgue assumption, \eqref{eq:V} gives that $V(\epsilon) \propto \epsilon^D$.
Hence the time per sample is asymptotic to the following expressions, written in terms of $\tau = 1/\epsilon$ for interpretability:
$C_1 = \tau^D$ for ABC and $C_2 = D^2 [\log \tau]^2 = [\log C_1]^2$ for RE-ABC.
Hence ABC has an exponential cost in $D$, while RE-ABC has only a quadratic cost.
This makes high-dimensional inference more tractable for RE-ABC but dimension reduction via summary statistics will remain useful in controlling the cost when $D$ is large.

These results assume the algorithms are run sequentially.
The PMMH stage of RE-ABC is innately sequential, but particle updates can be run in parallel, providing a benefit from parallelisation.
Compared to the most efficient ABC algorithms,
this is an advantage over ABC-MCMC and seems roughly comparable to that of ABC-SMC algorithms.

\section{Gaussian example} \label{sec:Gaussian}

In this section we compare ABC (Algorithm \ref{alg:ABCrs}) and RE-ABC (Algorithm \ref{alg:pmmh}) on a simple Gaussian model.
The model is $Y_i \sim N(0, \sigma^2)$ independently for $1 \leq i \leq 25$.
We use the prior $\sigma \sim \text{Uniform}(0, 10)$.
This is an interesting test case because $\dim(y)$ is large enough to cause difficulties for ABC methods but calculations are quick, and the results can be compared to those of likelihood-based methods.

\subsection{Comparison of ABC and RE-ABC}

We compared ABC and RE-ABC for observations drawn from the model using $\sigma=3$.
For each of $\epsilon=8,9,\ldots,30$, we ran ABC until $N=500$ simulations were accepted and calculated the root mean squared error and time per acceptance.
Both FIXED-RE-ABC and ADAPT-RE-ABC were run for 2000 iterations with $\epsilon=3, 5, 10, 15, 20, 25$.
As described in Section \ref{sec:inference}, pilot runs were used to tune the number of particles, the Metropolis-Hastings proposal standard deviation and, where necessary, the threshold sequence.
We chose the number of acceptances in all ADAPT-RE-ABC analyses to be half the number of particles.
To avoid dealing with burn-in, we started the PMMH chains at $\sigma=3$.
For comparison we also ran ABC-MCMC \citep{Marjoram:2003} and MCMC using the exact likelihood.

\begin{figure*}[pht]
\includegraphics[width=\textwidth]{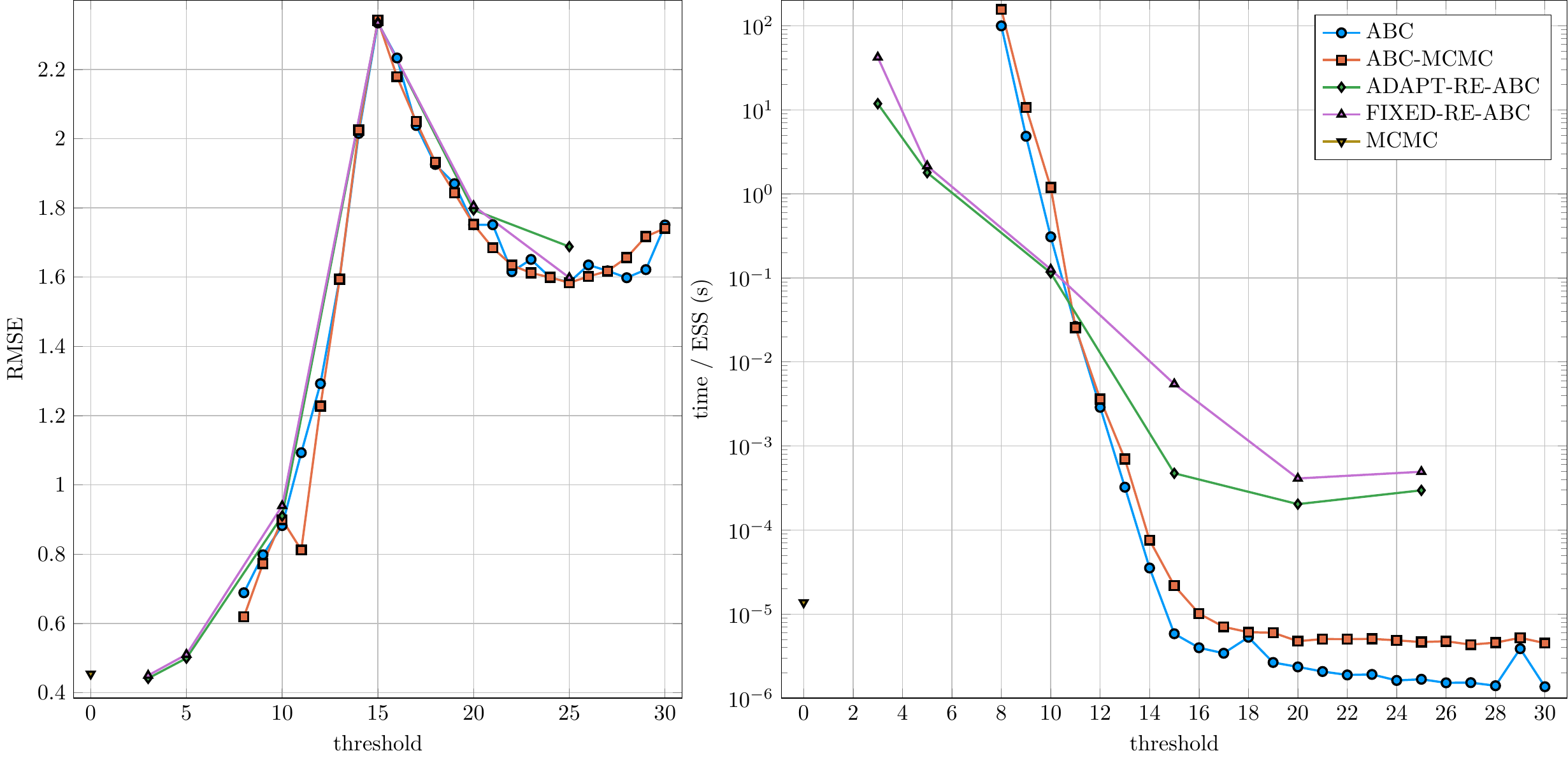}
\caption{Simulation study comparing ABC, ABC-MCMC, FIXED-RE-ABC, ADAPT-RE-ABC and exact likelihood MCMC on IID Gaussian data.} \label{fig:iidnormal}
\end{figure*}

Figure \ref{fig:iidnormal} shows the results.
The left panel illustrates that accuracy improves as the acceptance threshold $\epsilon$ is reduced below roughly 15, and, as expected, all methods produce very similar results.
In particular the biased likelihood estimates in ADAPT-RE-ABC have a negligible effect overall.
The right panel investigates the time taken per sample by ABC.
For MCMC output, this is time divided by the effective sample size (the IMSE estimate of \citealp{Geyer:1992}.)
Under ABC and ABC-MCMC, time per sample increases rapidly as $\epsilon$ is reduced.
For both RE-ABC algorithms the increase is slower, allowing smaller values of $\epsilon$ to be investigated.
Neither RE-ABC algorithm is obviously more efficient than the other.
This difference between ABC and RE-ABC is consistent with the asymptotics on computational cost described in Section \ref{sec:cost summary}.
However for large $\epsilon$ values ABC and ABC-MCMC are cheaper.
Overall RE-ABC permits smaller $\epsilon$ values to be investigated at a reasonable computational cost, producing more accurate approximations.

\begin{figure}[pht]
\begin{center}
\includegraphics[width=0.5\textwidth]{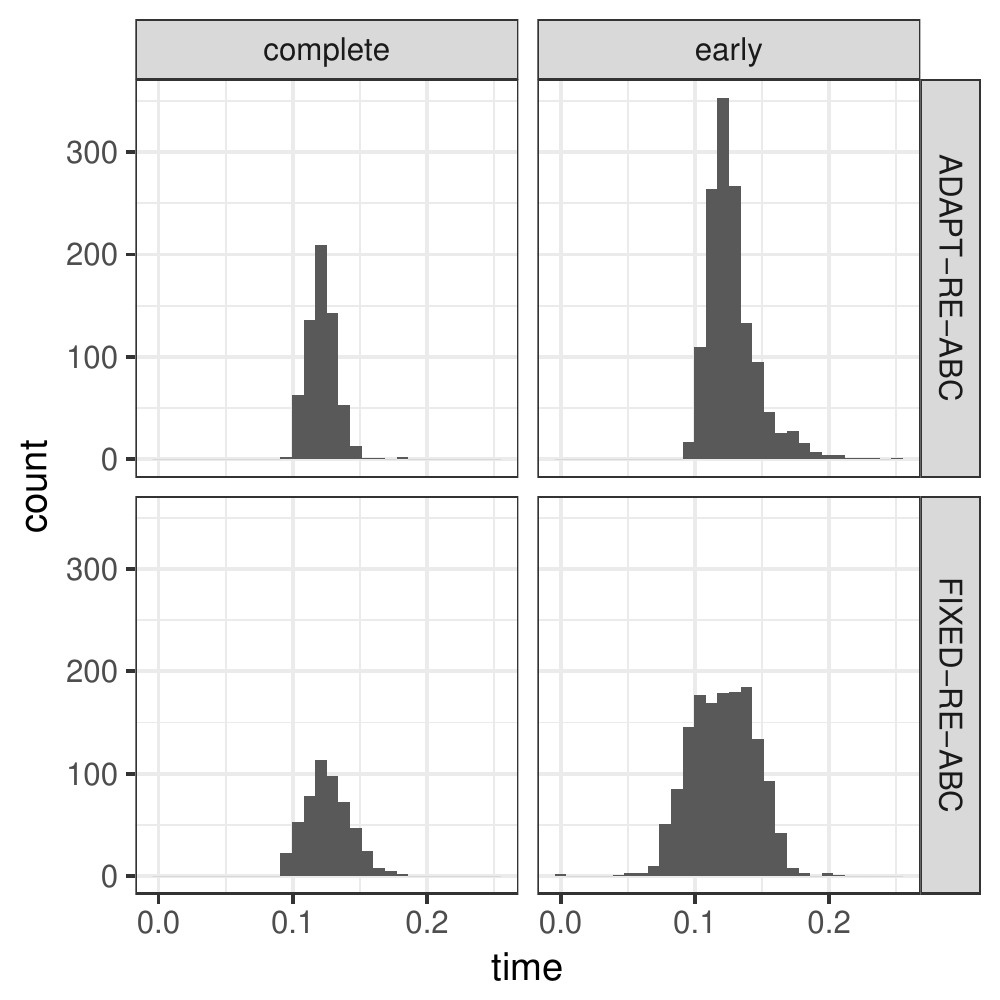}
\end{center}
\caption{Histograms of times (in seconds) taken by calls to RE-SMC within FIXED-RE-ABC and ADAPT-RE-ABC analyses of IID Gaussian data.
Both analyses used $\epsilon=5$ and the same tuning details, chosen using a pilot run.
The left column is for those calls in which RE-SMC was completed, while the right shows those where it was terminated early.
} \label{fig:times_normal}
\end{figure}

Figure \ref{fig:times_normal} provides some further insight into the efficiency of the RE-ABC algorithms, by looking at the times taken for calls to the RE-SMC algorithms.
These have similar distributions for FIXED-RE-ABC and ADAPT-RE-ABC, indicating that there is little difference in their efficiency.
One point of interest is that ADAPT-RE-SMC takes a minimum time of $0.095$ seconds even when it stops early, while FIXED-RE-SMC sometimes stops early in a much shorter time.
However this happens too rarely to have much effect on overall efficiency.

\subsection{Validity of assumptions} \label{sec:validation}

\begin{figure}[pht]
\begin{center}
\includegraphics[width=0.5\textwidth]{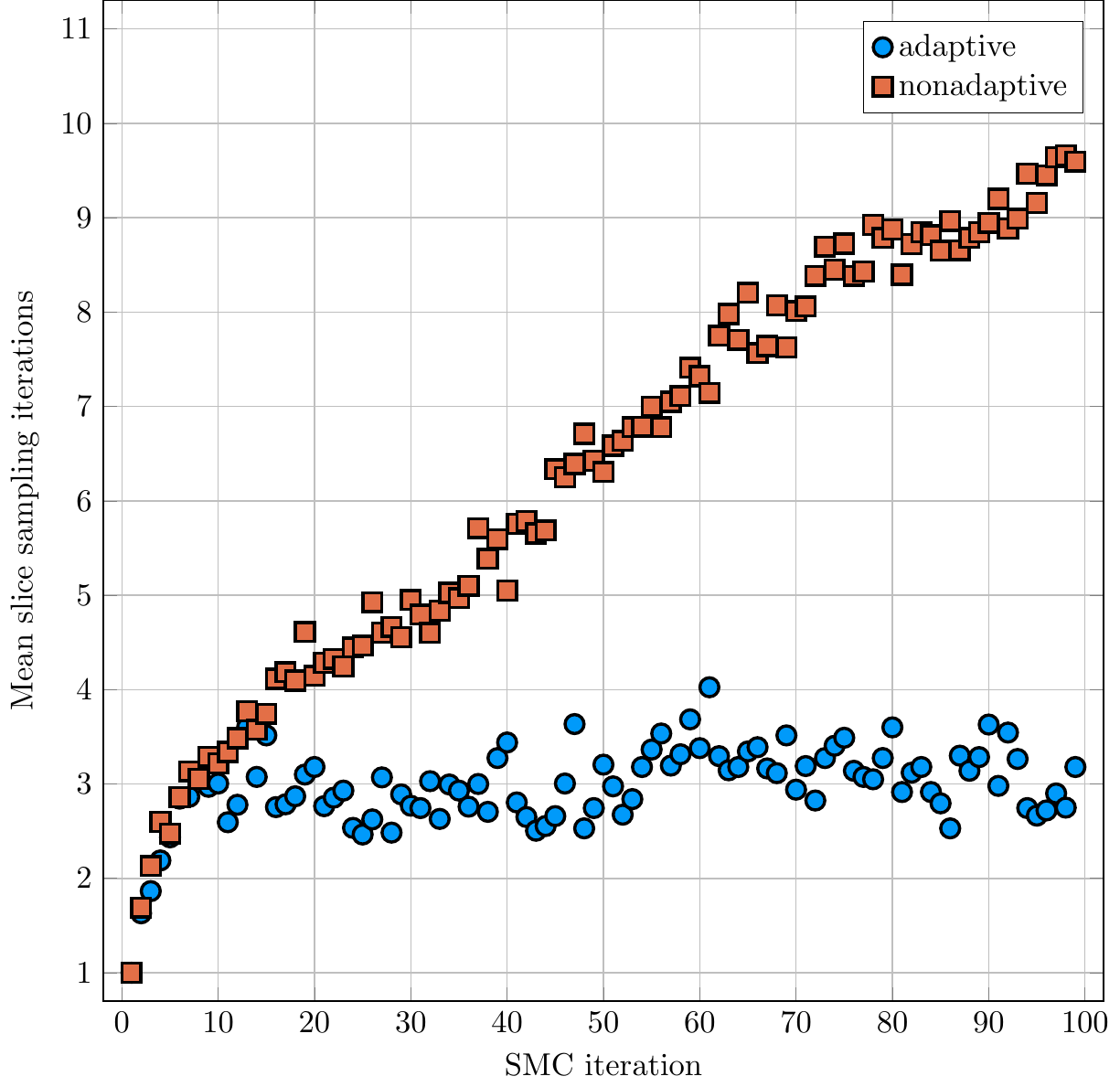}
\end{center}
\caption{Number of slice sampling iterations required under adaptive and non-adaptive rules for selecting the tuning parameter $w$ within a run of FIXED-RE-SMC on IID Gaussian data.} \label{fig:slice}
\end{figure}

We also used the Gaussian example to investigate the validity of various assumptions about RE-ABC used in this paper.
First we considered the cost of slice sampling calls in RE-SMC.
Figure \ref{fig:slice} shows the mean number of iterations that slice sampling requires during an illustrative FIXED-RE-SMC run.
Two cases are shown: non-adaptive slice sampling tuning ($w=1$ in Algorithm \ref{alg:slice}) or adaptive tuning ($w$ updated as described in Section \ref{sec:slice}).
This gives empirical evidence that adaptive tuning prevents the slice sampling cost from increasing during the algorithm, as desired.
Repeated trials show that both methods produce very similar mean likelihoods.
However adaptive tuning did increase the log-likelihood variance slightly so there is a small trade-off in its use.

\begin{figure}[pht]
\begin{center}
\includegraphics[width=0.5\textwidth]{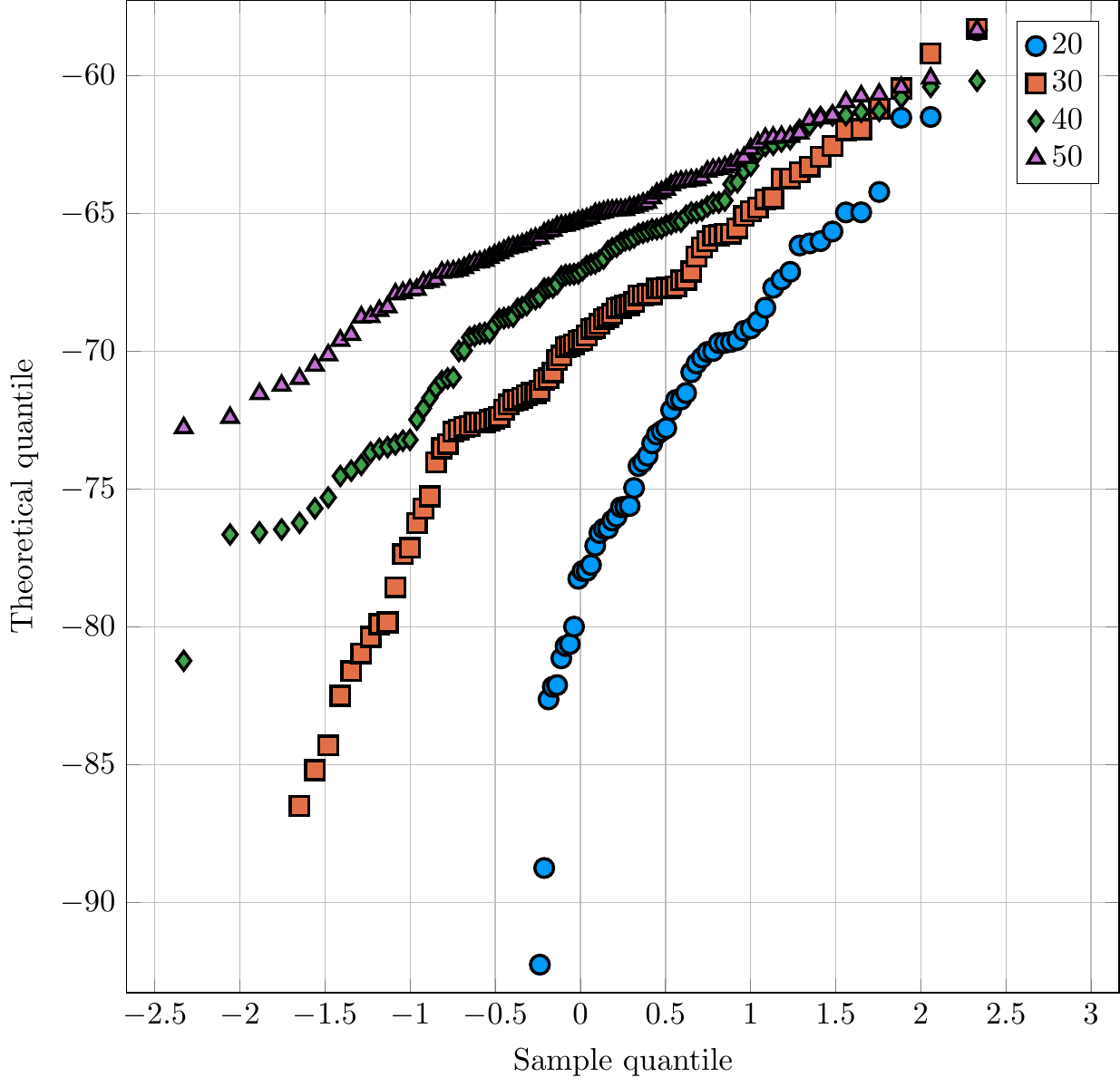}
\end{center}
\caption{Normal quantile-quantile plots of log likelihood estimates for an FIXED-RE-SMC example under various numbers of particles. Omitted points correspond to likelihood estimates of zero.} \label{fig:iidnormal_qq}
\end{figure}

Secondly we investigated the distribution of likelihood estimates produced by FIXED-RE-SMC given a particular $\theta$ value.
Recall that the theoretical literature on PMMH assumes that these follow a log-normal distribution.
Figure \ref{fig:iidnormal_qq} shows quantile-quantile plots comparing log likelihood estimates to normal quantiles.
The estimates are approximately normal when a sufficient number of particles are used, but become increasingly skewed as this shrinks.
A major departure from normality is that for a small number of particles many likelihood estimates are zero.
The corresponding points are omitted from the plot.
In conclusion, the normality assumption seems reasonable if a sufficient number of particles are used.

\section{Epidemic application} \label{sec:epidemics}

Infectious disease data is often modelled using \emph{compartment models}
where members of a population pass through several stages.
We will consider a model with susceptible, infectious and removed stages -- the so-called \emph{SIR model} \citep{Andersson:2000}.
A susceptible individual has not yet been infected with the disease but is vulnerable.
An infectious individual has been infected and may spread the disease to others.
A removed individual can no longer spread the disease.
Depending on the disease this may be due to immunity following recovery, or death.

We will use a stochastic version of this model based on a continuous time stochastic process $\{S(t), I(t): t \geq 0\}$ for numbers susceptible and infectious at time $t$.
The total population size is fixed at $n$ so the number removed at time $t$ can be derived as $R(t)=n-S(t)-I(t)$.
The initial conditions are $(S(0), I(0)) = (n-1, 1)$.
Two jump transitions are possible: infection $(i,j) \mapsto (i-1,j+1)$ and removal $(i,j) \mapsto (i,j-1)$.
The simplest version of the model is Markovian and is defined by the instantaneous hazard functions of the two transitions, which are $\frac{\lambda}{n} S(t) I(t)$ for infection and $\gamma I(t)$ for removal.
The unknown parameters are $\lambda$, controlling infection rates and $\gamma$, the removal rate.
A goal of inference is often to learn about the basic reproduction number $R_0 = \lambda / \gamma$.
This is the expected number of further infections caused by an initial infected individual in a large susceptible population.
When $R_0<1$, most epidemics will infect an insignificant proportion of a large population.
Many variations on the Markovian SIR model are possible, some of which are outlined below.

Likelihood-based inference is straightforward for fully observed data from an SIR model.
However in practice only partial and possibly noisy observations of removal times are available, producing an intractable likelihood.
For many models near-exact inference is possible by MCMC methods \citep[summarised by][]{McKinley:2014}, but small changes to the details require new and model-specific algorithms.
Approximate inference can be performed by ABC \citep[summarised by][]{Kypraios:2016}, which is more adaptable but does not scale well to high-dimensional data.
Here we illustrate how RE-ABC can, without modification, perform inference for several variations on the SIR model, and do so more efficiently than standard ABC methods.
As we concentrate on a classic and well-studied dataset, our analysis does not provide any novel subject-area insights.

Section \ref{sec:Sellke} describes a method of simulating from SIR models.
Section \ref{sec:SIRdist} discusses the distance function we use to implement RE-ABC.
Data analysis is performed in Section \ref{sec:Abakaliki}.

\subsection{Sellke construction} \label{sec:Sellke}

The Sellke construction \citep{Sellke:1983} for an SIR model provides an appealing way to simulate epidemic models.
It introduces latent \emph{infectious periods} $g_i \sim F_{\text{inf}}$ and \emph{pressure thresholds} $p_i \sim F_{\text{press}}$ for $1 \leq i \leq n$, all independent.
For the Markovian SIR model, $F_{\text{inf}}$ is $Exp(\gamma)$ and $F_{\text{press}}$ is $Exp(1)$, but other choices are possible and may be more biologically plausible.
We condition on $g_1=0$ so that the first infection occurs at time 0.
Algorithm \ref{alg:Sellke} shows how these variables and the parameter $\lambda$ are converted to simulated removal times.
To use slice sampling we require the latent variables to be uniformly distributed a priori.
Therefore we use quantiles of the $g_i$s and $p_i$s as the latent variables.

The cost of Algorithm \ref{alg:Sellke} is $O(n \log n)$, where $n$ is the population size.
This is because the main loop runs at most $2n-1$ times, and involves finding the minimum of a set of up to $n-1$ removal times, which requires $O(\log n)$ steps.
(This is the case if the set is stored as an ordered vector.
The cost of adding a new item is $O(\log n)$.)

Alternative simulation methods exist, principally the Gillespie algorithm \citep[described in][for example]{Kypraios:2016}.
Here the latent variables form a sequence controlling the behaviour of each successive jump event.
The Gillespie algorithm has the advantage of $O(n)$ cost.
However it seems hard for slice sampling to explore the space of latent variables
due to the behaviour of the mapping $y(\theta, x)$.
In particular a small change in latent variables which alters the type of one jump will typically have a large and unpredictable effect on all the subsequent jumps.
For more discussion on desirable properties of $y(\theta, x)$, see Section \ref{sec:conclusion}.

Note that when $F_{\text{press}}$ is $Exp(1)$ then $R_0 = \lambda E(F_{\text{inf}})$ \citep{Andersson:2000}.
However to our knowledge the definition of $R_0$ has not been extended to cover general $F_{\text{press}}$.

\begin{Algorithm}[htp]
\caption{Sellke construction epidemic simulator}
\label{alg:Sellke}
\begin{itemize}
\item[]
{\bf Input}: population size $n$, scaled infection rate parameter $\beta=\lambda/n$, infectious periods $g_1,\ldots,g_n$ and pressure thresholds $p_2,\ldots,p_n$.
\end{itemize}
\begin{enumerate}
\item Set $r_1 \mapsto g_1$ (assumes individual 1 has infection time 0).
\item Set $r_i \mapsto \infty$ for $i>2$.
\item Set $I \mapsto 1$ (current number infected), $t \mapsto 0$ (current time), $p \mapsto 0$ (current pressure).
\item {\bf While $I>0$:}
\item Find $p_a = \min \{ p_i | p_i>p \}$. If this set is empty use $p_{n+1} = \infty$.
\item Find $r_b = \min \{ r_i | r_i>t \}$.
\item Set $p' \mapsto p + \beta I (r_b-t)$ (pressure at time $r_b$ if $I$ does not change)
\item If $p_a < p'$:
\begin{enumerate}
  \item Set $I \mapsto I+1$, $t \mapsto t + \frac{p_a-p}{\beta I}$, $r_a \mapsto t + g_a$, $p \mapsto p_a$.
\end{enumerate}
\item Else:
\begin{enumerate}
  \item Set $I \mapsto I-1$, $t \mapsto r_b$, $p \mapsto p'$.
\end{enumerate}
\item {\bf End while}
\end{enumerate}
\begin{itemize}
\item[]
Output: Removal times $r_1,r_2,\ldots,r_n$.
Infinite removal time represents an individual who is never infected.
\end{itemize}
\end{Algorithm}

\subsection{Distance function} \label{sec:SIRdist}

Recall that the data are the inter-removal times, or equivalently the times since the first removal.
For a simulated dataset, let $r_{(1)} \leq r_{(2)} \leq \ldots \leq r_{(\nu)}$ denote the ordered removal times of a dataset with $\nu$ removals.
The times since first removal are then $s_{(i)} = r_{(i)} - r_{(1)}$ for $1 \leq i \leq \nu$.
Similar notation, with the addition of a subscript $\texttt{\text{obs}}$ will be used for the observed dataset.
We define the distance between a simulated and observed dataset as:
\begin{equation} \label{eq:SIRdist}
\begin{split}
&\left[ \sum_{i \leq \min(\nu_{\text{obs}}, \nu)} (s_{\text{obs}, (i)} - s_{(i)})^2 \right]^{1/2} \\
& + \sum_{\nu_{\text{obs}} < i \leq \nu} [k + \bar{\rho} - \rho_{(i)}] +
\sum_{\nu < i \leq \nu_{\text{obs}}} [k + \rho_{(i)}].
\end{split}
\end{equation}
Here $k$ is a tuning parameter penalising mismatches between $\nu$ and $\nu_{\text{obs}}$.
We take $k=1000$.
The $\rho_{(i)}$ terms are the sorted simulated pressure thresholds and $\bar{\rho}$ is the total simulated pressure (which equals $\beta$ times the sum of the infectious periods for removed individuals).
They are included to encourage these pressures to increase or decreasing appropriately to match $\nu$ and $\nu_{\text{obs}}$.
Without the pressure terms RE-SMC performed poorly due to the discrete nature of $\nu$.
See Section \ref{sec:extensions} for further discussion.

\subsection{Analysis of Abakaliki data} \label{sec:Abakaliki}

The Abakaliki dataset contains times between removals from a smallpox epidemic in which 30 individuals were infected from a closed population of 120.
It has been studied by many authors under many variations to the basic SIR model.
We study three models.
The first model uses a Gamma$(k, \gamma)$ infectious period (similar to \citealp{Neal:2005}).
The second assumes pressure thresholds are distributed by a Weibull$(k,1)$ distribution (as in \citealp{Streftaris:2012}.)
The third is the Markovian SIR model, but with removal times only recorded within 5 day bins.
This is realised by altering the $s_{\text{obs}, (i)} - s_{(i)}$ term (difference between simulated and observed day of removal) in \eqref{eq:SIRdist} to $f(s_{\text{obs}, (i)}) - f(s_{(i)})$ where $f(s) = 5\lfloor s/5 \rfloor$, the greatest multiple of 5 less than or equal to $s$.
In each model there are two or three unknown parameters:
$\lambda$, controlling infection rates;
$\gamma$, infectious period scale;
$k$, a shape parameter.
These are all assigned independent exponential prior distributions with rate 0.1, representing weakly informative prior beliefs that these parameters are less likely to be large.

We chose the acceptance threshold to be $\epsilon=15$ on the pragmatic grounds that this produced run-times of no more than 6 hours on a desktop PC.
Tuning was performed using pilot runs as described in Section \ref{sec:inference}.
Of particular note is the number of particles required: 300 (Gamma infectious period), 200 (Weibull pressure thresholds) and 400 (binned removal times).
First we present results for FIXED-RE-ABC, with discussion on ADAPT-RE-ABC to follow shortly.
Table \ref{tab:Abakaliki} summarises the approximate posterior results.
As the parameters differ between models, we don't present parameter estimates.
Instead we give several quantities of interest for each:
the $R_0$ estimate (where defined) and the means and standard deviations of (a) the pressure thresholds and (b) the infectious period.
Most quantities are similar to each other and previous analyses (see \citealp{McKinley:2014} for a summary of many of these) despite the different modelling assumptions.
A noticeable difference is that the infectious period is less variable in the model where it follows a Gamma distribution.

Figure 1 of the supplementary material shows simulated epidemics from each model.
This shows that our choice of $\epsilon$ produces epidemics reasonably close to the observed data for every model.
Formal model choice is not straightforward in our framework (see discussion in Section \ref{sec:conclusion}), but it is easy to explore whether the models produced large differences in log-likelihood.
In this case differences were modest, as shown by Figure 2 in the supplementary material, and within what would be explained, using BIC type arguments, by the differing number of parameters in the models.
So we conclude qualitatively that are no clear differences in fit between the models.

\begin{table*}[pht] \footnotesize
\begin{center}
\begin{tabular}{c|ccccc}
                            &             & \multicolumn{2}{c}{Pressure thresholds} & \multicolumn{2}{c}{Infectious period}                \\
Model                       & $R_0$       & Mean                                    & Standard deviation & Mean       & Standard deviation \\
\hline
5 day bins                  & 1.16 (0.30) & 0.11 (0.03)                             & 0.11 (0.03)        & 11.1 (3.0) & 11.1 (3.0)         \\
Gamma infectious period     & 1.18 (0.24) & 0.09 (0.03)                             & 0.09 (0.03)        & 13.6 (3.8) & 6.8 (2.2)          \\
Weibull pressure thresholds & -- 
                                          & 0.10 (0.04)                             & 0.11 (0.03)        & 12.4 (3.3) & 12.4 (3.3)
\end{tabular}
\end{center}
\caption{Approximate posterior estimates of basic reproduction number $R_0$ and the means and standard deviations of pressure thresholds and infectious periods for the Abakaliki data under three models computed using FIXED-RE-ABC.
The table contains Monte Carlo estimates along with standard deviations in brackets.
The $R_0$ value is not given for the Weibull pressure threshold model as no definition is available for this model.}
\label{tab:Abakaliki}
\end{table*}

\begin{figure}[pht]
\begin{center}
\includegraphics[width=0.5\textwidth]{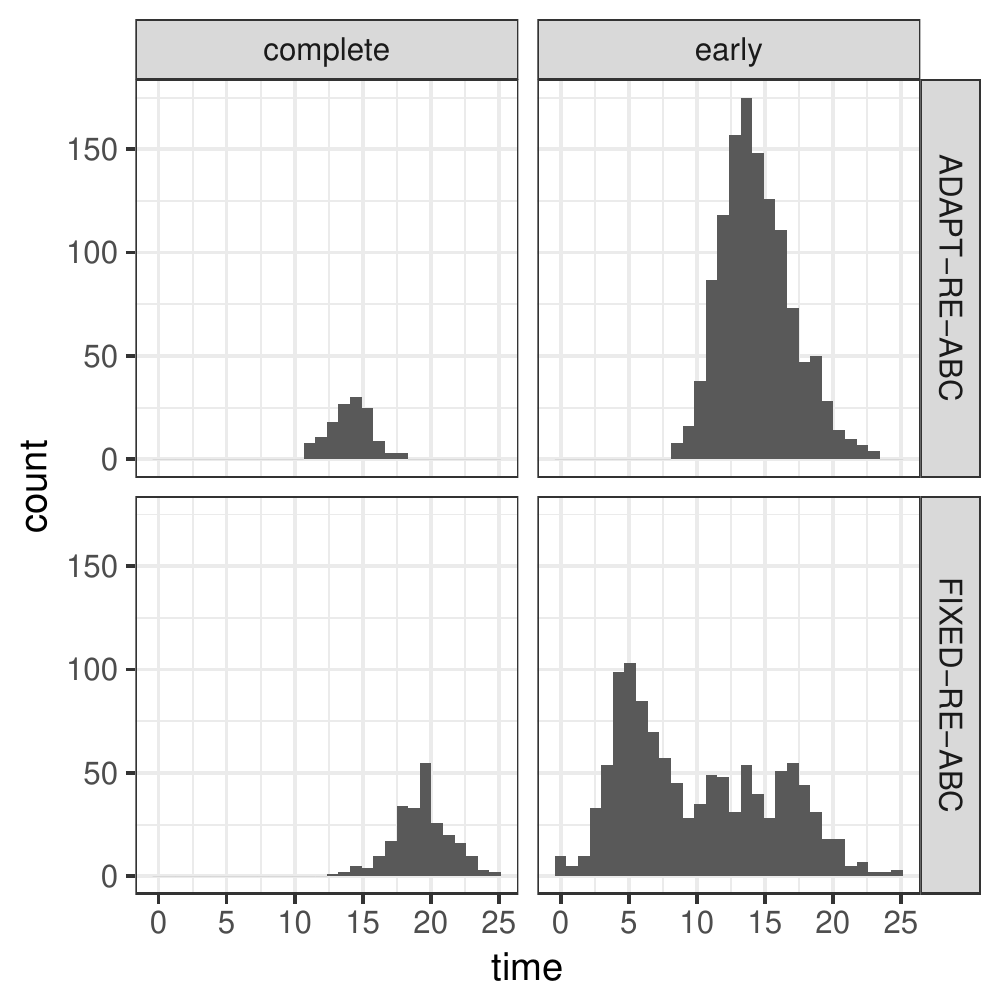}
\end{center}
\caption{Histograms of times (in seconds) taken by calls to RE-SMC within FIXED-RE-ABC and ADAPT-RE-ABC analyses of Abakaliki data.
Both analyses used $\epsilon=15$ and the same tuning details, chosen using a pilot run.
The left column is for those calls in which RE-SMC was completed, while the right shows those where it was terminated early.
} \label{fig:times_Abakaliki}
\end{figure}

ADAPT-RE-ABC was also tried and returned parameter inference results extremely similar to those for FIXED-RE-ABC -- see Table 1 in the supplementary material.
This shows that, as in Section \ref{sec:Gaussian}, the bias in its likelihood estimates has a negligible effect on the final results.
However, for some analyses the run times were longer.
For example, the Gamma infectious period model took 263 minutes for FIXED-RE-ABC and 323 minutes for ADAPT-RE-ABC.
Figure \ref{fig:times_Abakaliki} investigates this in more detail.
It shows that the run time difference is because most calls to RE-SMC terminate early, and these are generally quicker under FIXED-RE-SMC.
It is also interesting that ADAPT-RE-SMC is typically faster for completed RE-SMC calls.
These findings are discussed in the next section.

We also ran ABC-MCMC for comparison, using the same MCMC and $\epsilon$ tuning choices as for RE-ABC.
For run-times of comparable length to RE-ABC, ABC-MCMC produced too few acceptances to calculate effective sample sizes accurately.
Instead we consider the time per acceptance.
For ABC-MCMC this was at least 12 minutes for all models.
For RE-ABC this value was always less than 2 minutes.

\section{Discussion} \label{sec:conclusion}

We have presented a method for approximate inference under an intractable likelihood when simulation of data is possible.
It uses the same posterior approximation as ABC, \eqref{eq:ABCposterior}, which is controlled by a tuning parameter $\epsilon$.
The advantage of our method is that smaller values of $\epsilon$ can be achieved for the same computational cost, resulting in more accurate inference.
We have shown this is the case through asymptotics (Section \ref{sec:cost summary}) and empirically (Sections \ref{sec:Gaussian} and \ref{sec:epidemics}.)
This increased accuracy allows higher dimensional data or summary statistics to be analysed in practice.

\subsection{Latent variable considerations} \label{sec:latent}

Our method represents the model of interest with latent variables $x$,
and uses SMC and slice sampling to search for promising $x$ values.
For this search strategy to work well it seems necessary that:
\begin{itemize}
\item Evaluating $y(\theta,x)$ is reasonably cheap.
\item Sets of the form $\{ x | d(\yobs,y(\theta,x)) \leq \epsilon \}$ are easy to explore using slice sampling.
This would be difficult for sets made up of many disconnected components, or which are lower dimensional manifolds.
Smoothness of $y$ to changes in $x$ will help meet this condition.
\end{itemize}
Furthermore, our current implementation requires that the number of latent variables is fixed.
However the method could be adapted to the case of a variable number by altering the slice sampling algorithm (see Section 4.2 of \citealp{Murray:2016}).

\subsection{Adaptive and non-adaptive algorithms} \label{sec:adaptation}

The RE-ABC algorithm can use RE-SMC with a fixed $\epsilon$ sequence (FIXED-RE-SMC) or one that is chosen adaptively (ADAPT-RE-SMC).
FIXED-RE-SMC provides unbiased estimates of the ABC likelihood, as required by the PMMH algorithm, while ADAPT-RE-SMC has a small bias.
In practice we observe very little difference in the posterior results between the two algorithms, suggesting that this bias has a negligible effect in practice.
We also note that, if desired, a bias correction approach from \cite{Cerou:2012} could be applied.

Nonetheless, we recommend using the FIXED-RE-SMC algorithm within RE-ABC
(together with a pilot run of ADAPT-RE-SMC to choose the $\epsilon$ sequence.)
The main reason is that it is faster to run in practice, as found in Section \ref{sec:epidemics}.
Figure \ref{fig:times_Abakaliki} shows that this is because FIXED-RE-SMC can terminate more quickly for poor proposed $\theta$ values.
Interestingly, in the iterations where early termination is not required ADAPT-RE-SMC is slightly quicker.
We speculate that this is because it often finds a shorter $\epsilon$ sequence.
Furthermore, the theory of \cite{Cerou:2012} suggests that ADAPT-RE-SMC produces less variable ABC likelihood estimates, which would improve PMMH efficiency.
Therefore there may be some scope for a more efficient RE-SMC algorithm which combines the best features of the adaptive and non-adaptive approaches.

\subsection{Possible extensions} \label{sec:extensions}

\paragraph{More efficient $\epsilon$ sequence adaptation}

ADAPT-RE-ABC adapts the $\epsilon$ sequence for each $\theta$ value separately.
One alternative is to instead update the sequence based on information from SMC runs at previous $\theta$ values used by PMMH.
This could be done using stochastic approximation \citep[see e.g.][]{Andrieu:2008, Garthwaite:2016}, with the aim of making the $\hat{P}_t$ values in Algorithm \ref{alg:FIXED-RE-SMC} as similar as possible -- which minimises asymptotic variance of the likelihood estimates, as discussed in Section \ref{sec:RE-SMC}.
The result would be an adaptive MCMC algorithm, and it may be theoretically challenging to prove it has desirable convergence properties \citep{Andrieu:2008}.

\paragraph{Joint exploration of $(\theta,x)$}

Many $\theta$ values proposed by RE-ABC are rejected after calculating an expensive likelihood estimate.
An appealing alternative is to update the parameters $\theta$ conditional on sampled $x$ values, for example through a Gibbs sampler with state $(\theta,x)$.
Unfortunately in exploratory analyses of such methods we found the $\theta$ updates generally did not mix well.
The reason is that $x$ is much more informative for $\theta$ than the observations $\yobs$.
This results in small $\theta$ moves compared to the posterior's scale.

Alternatively, one could consider nesting an SMC algorithm to explore $x$ within one to explore $\theta$, following \cite{Chopin:2013} and \cite{Crisan:2016}.
Exploring $\theta$ could proceed by reducing $\epsilon$ at each iteration.
This might avoid the time penalty of ADAPT-RE-SMC when used in PMMH, discussed in Section \ref{sec:adaptation}.

\paragraph{Discrete data}

RE-SMC can struggle if there is a discrete data variable $x^*$.
It can be hard for SMC to move from accepting a set of latent variables $A$ to another $A'$ in which the range of possible $x^*$ values is smaller, because $\Pr(x \in A' | x \in A, \theta)$ may be very small.
The issue is particularly obvious for ADAPT-RE-SMC as the $\epsilon$ sequence may fail to move below some threshold for a large number of iterations.
For FIXED-RE-SMC it would instead result in high-variance likelihood estimates.
In Section \ref{sec:SIRdist} this problem occurs for $\nu$, the number of removals.
There we adopt an application-specific solution by introducing continuous latent variables (pressure thresholds) into the distance function \eqref{eq:SIRdist}.
It would be useful to investigate more general solutions from the rare event literature (e.g.~\citealp{Walter:2015}).
Despite these potential issues, RE-ABC can perform well with discrete data in practice, for example in the binned data model of Section \ref{sec:Abakaliki}.

\paragraph{Non-uniform ABC kernels}

In this paper, the ABC likelihood \eqref{eq:ABClikelihood} is a convolution of the exact likelihood and a uniform kernel \eqref{eq:uniform kernel}.
Alternative kernel functions have also been used in ABC (e.g.~\citealp{Wilkinson:2013}) such as a Gaussian: $k(y;\epsilon) \propto \exp[-\tfrac{1}{2 \epsilon^2} d(y, \yobs)]$.
RE-ABC could easily be adapted to make use of these, but it is not clear what effect it would have on our asymptotic results.

\paragraph{Estimating log-likelihood gradients}

Where log-likelihood gradients can be estimated they allow more efficient inference schemes based on stochastic gradient descent \citep{Poyiadjis:2011} or MCMC \citep{Dahlin:2015}.
Estimating such gradients from SMC algorithms is possible using the Fisher identity \citep{Poyiadjis:2011}.
However the calculation would involve evaluating $\nabla_\theta y(\theta,x)$, which may be demanding for complicated $y$ functions.
\cite{Moreno:2016} use automatic differentiation to evaluate this for some models.
Alternatively, \cite{Andrieu:2012} propose using infinitesimal perturbation analysis methods.
It would be interesting to use either approach with RE-ABC.

\paragraph{Model choice}

A desirable extension to RE-ABC would be methods for model choice.
Possible methods to extend our PMMH approach include reversible jump MCMC or using a deviance information criterion.
See \cite{Chkrebtii:2015} and \cite{Francois:2011} for versions of these methods in the ABC context.
Alternatively, it may be more fruitful to use our likelihood estimate in algorithms which directly output model evidence estimates, such as importance sampling or population Monte Carlo \citep{Cappe:2004}.

\begin{acknowledgements}
We thank Chris Sherlock for suggesting the use of slice sampling and Andrew Golightly for helpful discussions.
\end{acknowledgements}

\appendix

\section{Computational cost} \label{sec:cost}

This appendix justifies the computational costs of ABC and RE-ABC stated in Section \ref{sec:cost summary}.
The argument for ABC is rigorous, while that for RE-ABC is more heuristic.
Note that throughout this appendix there is no need to distinguish between FIXED-RE-ABC and ADAPT-RE-ABC.

The results are for the asymptotic regime of small $\epsilon$ and hold for almost all $\yobs$.
We make several assumptions:
\begin{itemize}
\item[A1] The density $\pi(y|\theta)$ is with respect to Lebesgue measure $dy$ of dimension $D$.
\item[A2] The distance function is Euclidean distance.
\item[A3] Running slice sampling once requires $O(1)$ function evaluations.
\item[A4] RE-SMC uses $O \left( -\log \Pr(d(y,\yobs) \leq \epsilon | \theta) \right)$ iterations.
\item[A5] The time required to evaluate $y(\theta,x)$ is bounded above and below by non-zero constants which do not depend on $\theta$ or $x$.
\end{itemize}
Also, we will usually focus on the case where $D$ is asymptotically large.

Informally, A1 requires that all components of $y$ have continuous distributions.
Under A2 a key mathematical result below, \eqref{eq:LDT}, follows easily.
Also, a consequence of A2 which we will use is that, from \eqref{eq:V}, $V(\epsilon) \propto \epsilon^{-D}$.
A3 states that the cost of slice sampling does not increase as $\epsilon$ shrinks.
This is plausible due to our adaptive choice of $w$ (see Section \ref{sec:slice}), and was empirically verified above (see Figure \ref{fig:slice}.)
It follows that running RE-SMC requires $O(NT)$ function evaluations:
the number is asymptotic to the number of particles multiplied by the number of SMC iterations.
A4 states that the number of iterations used by RE-SMC is asymptotically proportional to the log of the rare probability being estimated.
This follows from a result of \cite{Cerou:2012}, reviewed in Section \ref{sec:RE-SMC},
that when the RE-SMC algorithm is tuned optimally
$\Pr(A_{k+1} | \theta, x \in A_k)$ is constant, say $\alpha$, where $A_k$ denotes the event $d(y(\theta,x),\yobs) \leq \epsilon_k$.
Therefore $\Pr(d(y,\yobs) \leq \epsilon | \theta) = \alpha^T$, and taking logs gives A4.
So the assumption is that RE-SMC is tuned to perform similarly to optimal tuning.
Assumption A5 states that performing a simulation has a minimum and maximum time requirement regardless of the inputs, which is usually reasonable.
This ensures that computation time is asymptotic to the number of simulations performed.

Many of these assumptions can be weakened.
This is discussed in the supplementary material, especially for the case of the epidemic model of Section \ref{sec:epidemics}.

\subsection{ABC} \label{sec:ABCcost}

Consider the probability of a simulation being accepted given $\theta$:
\[
{\Pr}(d(y, \yobs) \leq \epsilon | \theta) = \int \pi(y|\theta) \mathbbm{1}(d(y, \yobs) \leq \epsilon) dy.
\]
By the Lebesgue differentiation theorem (see \citealp{Stein:2009} for example) for almost all $\yobs$:
\begin{equation} \label{eq:LDT}
\lim_{\epsilon \to 0} V(\epsilon)^{-1} \int \pi(y|\theta) \mathbbm{1}(d(y, \yobs) \leq \epsilon) dy = \pi(\yobs | \theta),
\end{equation}
Hence for small $\epsilon$:
\begin{equation} \label{eq:accprob}
\Pr(d(y, \yobs) \leq \epsilon | \theta) \sim V(\epsilon),
\end{equation}
where $\sim$ represents an asymptotic relation.
(Note that while $\pi(\yobs | \theta)$ does not affect this asymptotic relationship,
the acceptance probability will decrease for small $\pi(\yobs | \theta)$ i.e.~for poor $\theta$ choices.)

By assumption A5 the time per accepted sample is asymptotic to the number of simulations per accepted sample.
Using \eqref{eq:accprob}, the latter is asymptotic to $1/V(\epsilon)$.
In the case of large $D$ assumption A2 gives that this is $O(\tau^D)$, where $\tau = 1/\epsilon$.
For ABC versions of MCMC and SMC, time per accepted sample (or effective sample) is also bounded below by $\min_{\theta} \Pr(d(y,\yobs) \leq \epsilon | \theta)^{-1}$, so the same result applies.

\subsection{RE-ABC} \label{sec:RE-ABC cost}

For simplicity we analyse RE-ABC without the possibility of early termination in the RE-SMC algorithm.
An algorithm including early termination will give the same output for a smaller computational cost, although we suspect the gain is only likely to be a $O(1)$ factor.
Using the asymptotic results reviewed in Section \ref{sec:background} on SMC likelihood estimation and PMMH we conclude the following.
The number of particles in RE-SMC should be $N=O(T)$ to give a likelihood estimator whose log has variance $O(1)$, which optimises efficiency when these estimates are used in PMMH.
So, using A3, the number of simulations required by an iteration of RE-ABC is $O(T^2)$.
Using A4 and \eqref{eq:accprob} gives $T = O(-\log V(\epsilon))$.

So the number of simulations required per iteration of RE-ABC is $O([\log V(\epsilon)]^2)$.
In the case of large $D$ using A2 gives that this is  $O(D^2 [\log \tau]^2)$.
As in the previous section, assumption A5 implies these expressions also give the time per sample of RE-ABC.
They are also valid for the more relevant quantity of time per effective sample since effective sample size is proportional to the actual sample size.

\bibliography{highdimABC}

\end{document}


\maketitle

\section{Further results from Abakaliki example}

\begin{figure*}[pht]
\begin{center}
\includegraphics[width=\textwidth]{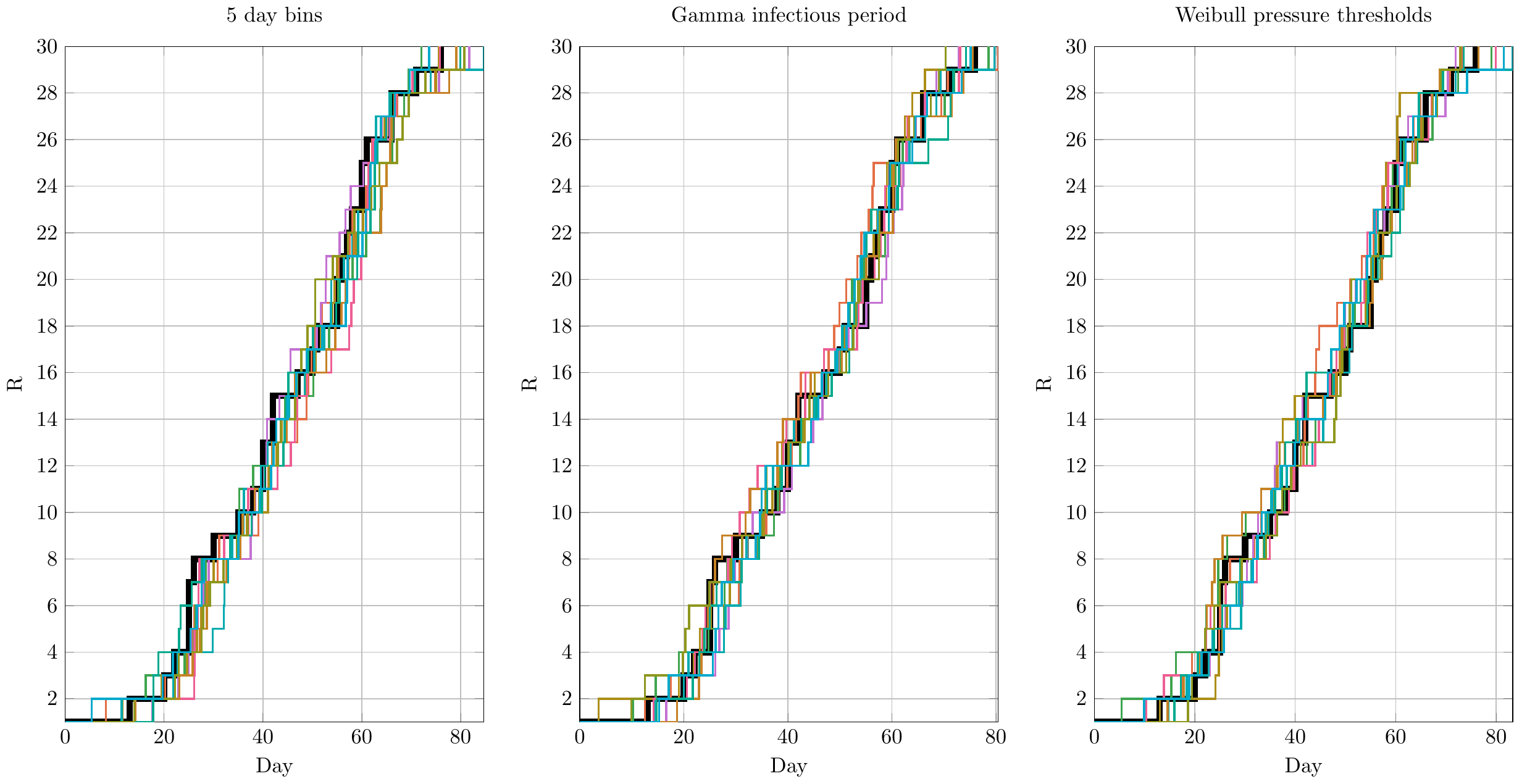}
\end{center}
\caption{Observed data (black line) and accepted simulations under FIXED-RE-ABC (coloured lines) for an analysis of Abakaliki data.
The x-axis shows number of days since the first removal, and the y-axis shows the total number of removals so far.
Accepted simulations are generated by running FIXED-RE-ABC for 10 parameter values taken from thinned MCMC output and selecting one particle from the final iteration whose distance is below the threshold $\epsilon$.
} \label{fig:Abakaliki sims}
\end{figure*}

\begin{figure}[pht]
\begin{center}
\includegraphics[width=0.5\textwidth]{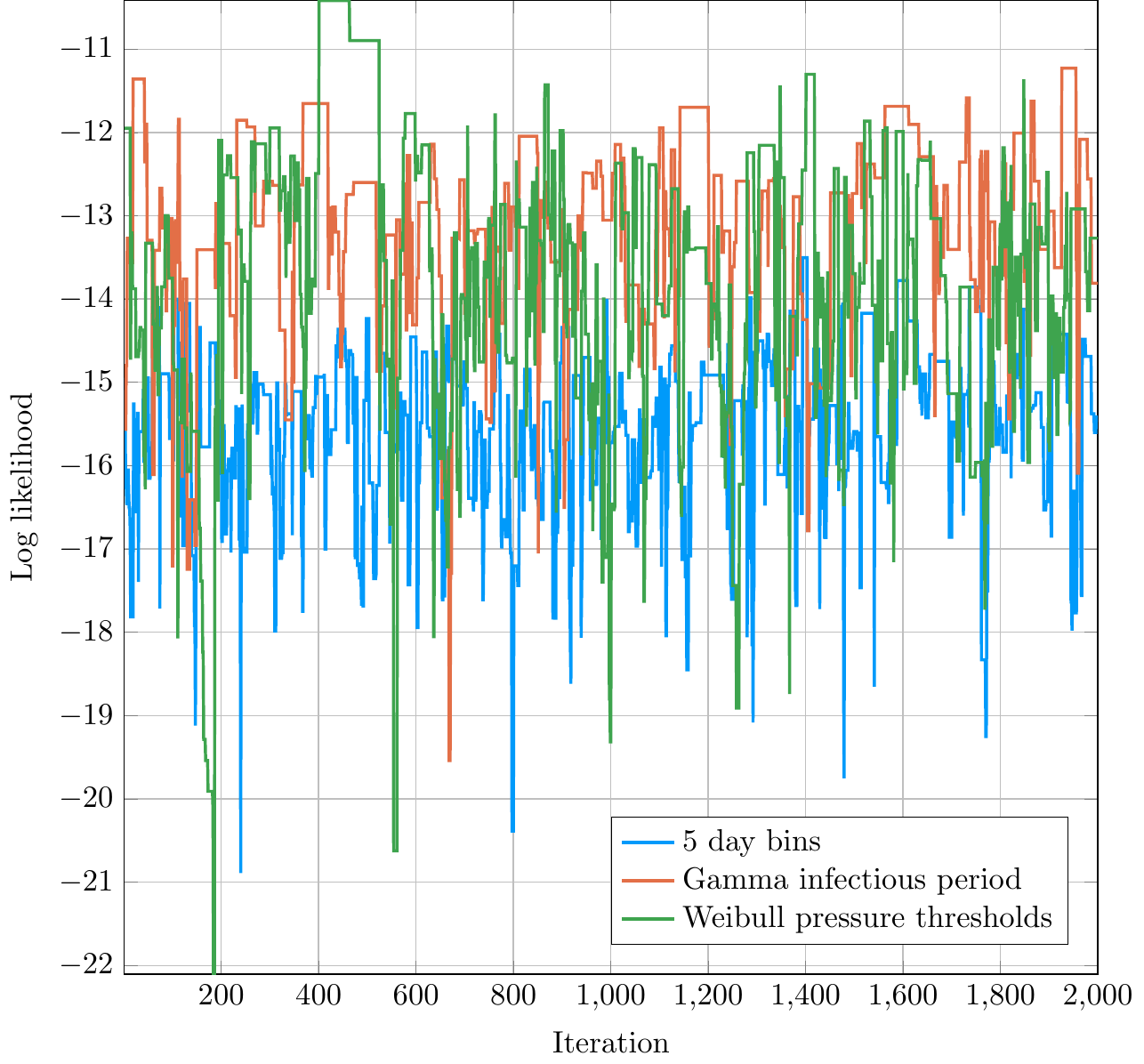}
\end{center}
\caption{Trace plots of log likelihood estimates resulting from FIXED-RE-ABC analyses of Abakaliki data.}
\label{fig:ll}
\end{figure}

This section reports some further details of our analysis of the Abakaliki data.
Table \ref{tab:Abakaliki adaptive} contains parameter estimates from ADAPT-RE-ABC analyses.
Figure \ref{fig:Abakaliki sims} shows simulated epidemics from each model using FIXED-RE-ABC.
Figure \ref{fig:ll} shows trace plots of log-likelihood estimates produced by FIXED-RE-ABC.
These are all discussed in the main text.

\begin{table*}[pht] \scriptsize
\begin{center}
\begin{tabular}{c|ccccc}
                            &             & \multicolumn{2}{c}{Pressure thresholds} & \multicolumn{2}{c}{Infectious period}                \\
Model                       & $R_0$       & Mean                                    & Standard deviation & Mean       & Standard deviation \\
\hline
5 day bins                  & 1.17 (0.29) & 0.11 (0.03)                             & 0.11 (0.03)        & 11.4 (2.9) & 11.4 (2.9)         \\
Gamma infectious period     & 1.16 (0.22) & 0.08 (0.03)                             & 0.08 (0.03)        & 14.9 (4.2) & 6.2 (1.8)          \\
Weibull pressure thresholds & - 
                                          & 0.10 (0.04)                             & 0.11 (0.04)        & 12.2 (3.8) & 12.2 (3.8)
\end{tabular}
\end{center}
\caption{ADAPT-RE-ABC posterior estimates of basic reproduction number $R_0$ and the means and standard deviations of pressure thresholds and infectious periods for the Abakaliki data under three models.
Monte Carlo estimates are quoted along with standard deviations in brackets.
The $R_0$ value is not given for the Weibull pressure threshold model as no definition is available for this model.
}
\label{tab:Abakaliki adaptive}
\end{table*}

\newpage

\section{Asymptotics with weaker assumptions}

This section discusses extending the asymptotic theory of the main paper to use weaker assumptions,
in particular showing how it can be used with most of the SIR models from the main paper.
For these models assumption A1 (data has density with respect to Lebesgue measure) does not hold because the data is not continuous, instead involving a discrete observation of the number of removals, $\nu$, and $\nu-1$ continuous inter-removal times.
Furthermore assumption A2 (distance function is Euclidean) does not hold since a more complicated distance function was used.
Our argument can be adapted to this model by showing that both assumptions effectively hold for sufficiently small $\epsilon$ values.
This is discussed in Section \ref{sec:A1A2}.

A further problem arises because the observed data contains repeated recovery times.
This means the data is on the boundary of the model's support, which causes technical problems.
While the asymptotic results of the main paper remain true for almost all $\yobs$ values, they are not necessarily valid when $\yobs$ is on this boundary.
This problem is discussed in Section \ref{sec:boundary}.

Finally, note that the main paper includes a model with discrete summary statistics: days of removal rounded down to a multiple of 5.
Here a sufficiently small non-zero $\epsilon$ value ensures an exact match of simulated and observed data.
Therefore it is not of interest to consider small $\epsilon$ asymptotics for this case.

\subsection{Weakening assumptions A1 and A2} \label{sec:A1A2}

Suppose that assumptions A1 and A2 do not hold, but there is some $\epsilon_0>0$ with the following properties.

\begin{itemize}
\item[B1] There is an injective mapping $z$ from $A=\{ y | d(y,\yobs) < \epsilon_0 \}$ to $\mathbb{R}^{D'}$.
\item[B2] The distribution $z(y) |\theta, y \in A$ has density $\pi(z|\theta)$ with respect to Lebesgue measure $dy$ of dimension $D'$.
\item[B3] For $y \in A$, $d(y, \yobs)$ equals $d_E(z(y), \zobs)$ where $d_E$ denotes Euclidean distance and $\zobs=z(\yobs)$.
\end{itemize}

\subsubsection{SIR model}

For the SIR model we can select $\epsilon_0$ such that $d(y,\yobs) < \epsilon_0$ guarantees that $y$ has the same number of removals as $\yobs$.
For example $\epsilon_0 = k$ will achieve this (recall that $k$ is a penalty in the distance function for the wrong number of removals).
Suppose $y$ has removal times $r_{(1)} \leq r_{(2)} \leq \ldots \leq r_{(\nu)}$ where $\nu$ is the number of removal times.
Let $z(y)$ be the dimension $\nu-1$ vector of times since first removal i.e.~the values $s_{(i)}=r_{(i)} - r_{(1)}$ for $2 \leq i \leq \nu$.
This meets the assumptions B1-B3.

\subsubsection{Asymptotics}

For both ABC and RE-ABC the crucial quantity is $\Pr(d(y,\yobs) \leq \epsilon | \theta)$.
For $\epsilon<\epsilon_0$ this is given by:
\[
\Pr(y \in A | \theta) \Pr(d(y,\yobs) \leq \epsilon | \theta, y \in A)
\]
Let these probabilities be $P_1$ and $P_2$ respectively.
Since the former does not depend on $\epsilon$ we have $P_1=O(1)$.
The latter is
\begin{align*}
\Pr(d_E(z(y),\zobs) \leq \epsilon | \theta, y \in A)
= \int \pi(z|\theta) \mathbbm{1}(d_E(z,\zobs) \leq \epsilon) dz
\end{align*}
Now we can repeat the argument of the main paper.
By the Lebesgue differentiation theorem for almost all $\zobs$:
\[
\lim_{\epsilon \to 0} \frac{\int \pi(z|\theta) \mathbbm{1}(d(z, z_\text{obs}) \leq \epsilon) dz}{\int \mathbbm{1}(d(z, z_\text{obs}) \leq \epsilon) dz} = \pi(z_\text{obs} | \theta).
\]
Hence for small $\epsilon$,
\[
P_2 \sim \int \mathbbm{1}(d(z, z_\text{obs}) \leq \epsilon) dz = O(\epsilon^{D'}).
\]
Using the arguments in the main paper it follows that the time per sample in ABC is $(P_1 P_2)^{-1}$ which is $O(\epsilon^{-D'})$,
and the time per effective sample for RE-ABC is $O({D'}^2 [\log \epsilon]^2)$.

\subsection{Data in the boundary of the support} \label{sec:boundary}

Our asymptotics rely on the Lebesgue differentiation theorem which states that when $g(y)$ is Lebesgue integrable and $dy$ is Lebesgue measure then the following holds for almost all $y_0$:
\begin{equation} \label{eq:LDT}
\lim_{\epsilon \to 0} \frac{\int g(y) \mathbbm{1}(y \in B_\epsilon) dy}{\int \mathbbm{1}(y \in B_\epsilon) dy} = g(y_0),
\end{equation}
where $B_\epsilon$ is a ball of radius $\epsilon$ centred on $y_0$.

However this is not true for $y_0$ on the boundary of the support of $g(y)$.
For example suppose $g(y)$ is a uniform density on $[0,1]$ and $y_0=0$.
Then for $\epsilon < 1$:
\[
\frac{\int g(y) \mathbbm{1}(y \in B_\epsilon) dy}{\int \mathbbm{1}(y \in B_\epsilon) dy} = 1/2
\]

This problem can be avoided by replacing $B_\epsilon$ with $B'_\epsilon = B_\epsilon \cap \text{supp}(g)$.
The Lebesgue differentiation theorem remains true in this case (see \citealp{Stein:2009}), as the $B'_\epsilon$ sets meet the condition of \emph{bounded eccentricity}.
That is, each $B'_\epsilon$ is contained in some ball $B$ such that $|B'_\epsilon| \leq \delta |B|$ for some constant $\delta$.
(This method could be also be used to show our asymptotics hold for many non-Euclidean distance functions.)

\bibliography{highdimABC}